\documentclass[final,1p,times]{elsarticle}

\biboptions{square,comma,numbers,sort&compress}

\usepackage{amsmath,amssymb,mathtools,bm}
\usepackage{caption}
\usepackage{hyperref,cleveref}
\hypersetup{pdfauthor={Kumpei Tsuji; Ciara Galvin; Hiroyuki Omura; Naoto Mitsume; Mitsuteru Asai}}
\usepackage{algpseudocode}
\usepackage{algorithm}
\usepackage{float}
\usepackage{placeins}
\usepackage{microtype}
\hypersetup{hidelinks}

\journal{Computational Particle Mechanics}

\begin{document}


\begin{frontmatter}

\title{Partitioned iterative FD-SPH for fluid–rigid-body interaction: Stability and validation for lightweight floating bodies} 


\author[label1]{Kumpei Tsuji\corref{cor1}}
\ead{kumpei.tsuji.e1@tohoku.ac.jp}

\author[label1]{Ciara Galvin}
\ead{galvin.ciara.s1@dc.tohoku.ac.jp}

\author[label2]{Hiroyuki Omura}
\ead{homura@bosai.go.jp}

\author[label3]{Naoto Mitsume}
\ead{mitsume@kz.tsukuba.ac.jp}

\author[label4]{Mitsuteru Asai\corref{cor2}}
\ead{asai@doc.kyushu-u.ac.jp}

\affiliation[label1]{organization={Department of Civil and Environmental Engineering, Tohoku University},
            addressline={Aramaki 6-6-06, Aza-Aoba, Aoba-ku},
            city={Sendai},
            postcode={980-8579},
            state={Miyagi},
            country={Japan}}

\affiliation[label2]{organization={National Research Institute for Earth Science and Disaster Resilience},
            addressline={Tennodai 3-1},
            city={Tsukuba},
            postcode={305-0006},
            state={Ibaraki},
            country={Japan}}

\affiliation[label3]{organization={Institute of Systems and Information Engineering, University of Tsukuba},
            addressline={Tennodai 1-1-1},
            city={Tsukuba},
            postcode={305-8573},
            state={Ibaraki},
            country={Japan}}

\affiliation[label4]{organization={Disaster Risk Reduction Center, Department of Civil Engineering, Kyushu University},
            addressline={Motooka 744, Nishi-ku},
            city={Fukuoka},
            postcode={819-0395},
            state={Fukuoka},
            country={Japan}}
        
\cortext[cor1]{Corresponding author. 
  Email: kumpei.tsuji.e1@tohoku.ac.jp, 
  Tel: +81-22-795-7418}

\cortext[cor2]{Corresponding author. 
  Email: asai@doc.kyushu-u.ac.jp, 
  Tel: +81-92-802-3373}



\begin{abstract}
This study develops and assesses a rigid-body-specific implementation of partitioned iterative fictitious-domain smoothed particle hydrodynamics (FD-SPH) for free-surface fluid--rigid-body interaction. Extending an existing strong-coupling framework for deformable structures, it combines volume-based hydrodynamic force and torque evaluation with Newton--Euler rigid-body updates. Within each time step, the mismatch between pressure-projected fictitious-fluid and rigid-body velocities is fed back through partitioned iterations. A fractional-step interpretation relates FD-SPH to the Passively Moving Solid (PMS) method and identifies conditions under which PMS projection and Newton--Euler updates yield identical rigid-body velocities. Single-pass and iterative FD-SPH are compared under matched numerical conditions. A two-dimensional settling-disk benchmark yields comparable responses for both schemes, consistent with published results. In three-dimensional buoyancy-driven cube tests, single-pass coupling becomes unstable at specific gravities of 0.1 and 0.2. At a specific gravity of 0.1, reducing the time-step size by a factor of 500 does not suppress the instability. Iterative strong coupling remains stable across the tested range of 0.1–1.0. Predicted equilibrium heights and orientations agree with hydrostatic expectations and published stable configurations, respectively. Two- and three-dimensional wet-bed dam-break simulations reproduce the principal experimental trends in wave elevation and floating-body heave and pitch. A simulation of two star-shaped rigid bodies demonstrates applicability to complex moving geometries without explicit interface reconstruction or surface-normal evaluation. All reported production strong-coupling simulations satisfy the prescribed relative-residual criterion at every physical time step before reaching the maximum coupling-iteration count. These findings support iterative FD-SPH for lightweight and complex floating bodies in transient free-surface flows.
\end{abstract}




\end{frontmatter}



\section{Introduction}
\label{sec1:Introduction}

Particle methods, including Smoothed Particle Hydrodynamics (SPH)~\cite{Lucy_SPH_1977,Gingold-Monaghan_SPH_1977} and the Moving Particle Semi-implicit (MPS) method~\cite{Koshizuka_MPS_1996,Koshizuka_MPS_book_2018}, provide mesh-free descriptions of moving boundaries and free-surface flows. Their geometric flexibility has supported applications to free-surface hydrodynamics~\cite{Asai_StabISPH_2012,Marrone_SPH-FSD_2010,Matsunaga_StabLSMPS_2022,Morikawa_ALE-ISPH_DPS_2023}, multiphase, phase-change, and seepage flows~\cite{Harada_MPS-DEM_morpho_2021,Wang_multi-phase_LSMPS_2022,Tsuji_ISPH-DEM_seepage_2023,Tsuji_SPH2_DPS_seepage_2025}, and fluid--structure interaction~\cite{Khayyer_ISPH-SPH_FSI_2018,Sun_FSI-SPH_2021,Lu_SPH_FSI_shell_2026}. Free-surface fluid--rigid-body interaction is particularly relevant to drifting debris during floods and tsunamis, floating structures, and offshore renewable-energy systems~\cite{Koshizuka_MPS_PMS_origin_1998,Liu_SPH_rigid_2014,Sun_SPH_FRI_2015,Asai-Li_ISPH_DEM_ETI_2021,Tan_SPH_wind-turbine_2023}. Reliable prediction of these interactions requires both an appropriate representation of the moving body and consistent coupling between the fluid and rigid-body motion.

Two representative approaches illustrate how rigid-body velocities are determined in particle-based fluid--rigid-body simulations~\cite{Negishi_MPH_PMS_2024}. In approaches based on explicit hydrodynamic loads, hydrodynamic force and torque are evaluated from fluid stresses or particle-based boundary forces and used to update rigid-body motion through the Newton--Euler equations~\cite{Asai-Li_ISPH_DEM_ETI_2021,Amaro_DEM-MPS_ice_2021,Li_MPS_floating-box_2025}. In contrast, the Passively Moving Solid (PMS) model proposed by Koshizuka et al.~\cite{Koshizuka_MPS_PMS_origin_1998} represents a rigid body using particles that first undergo a fluid-like calculation. Their velocities are subsequently corrected to satisfy rigid-body motion while preserving linear and angular momenta. This procedure avoids explicit surface-stress integration and surface-normal evaluation, and its simple implementation has supported applications to floating bodies and drifting objects~\cite{Wang_EMPS_PMS_DEM_2019,Negishi_MPH_PMS_2024}.

Beyond the determination of rigid-body motion, the temporal coupling procedure governs how the fluid and body solutions are reconciled within each physical time step. At the level of the coupled solution, monolithic methods solve the fluid and solid unknowns together, whereas partitioned methods retain separate solvers and exchange coupling data~\cite{Degroote_monolithic_partitioned_2009}. Probst and Teschner~\cite{Probst-Teschner_Monolithic-SPH_fluid-rigid_2023}, for example, developed a monolithic SPH formulation combining incompressibility, fluid--rigid-body coupling, and frictional contact. Partitioned methods provide an alternative that permits the reuse and independent development of fluid and rigid-body solvers. Relaxation and quasi-Newton techniques can accelerate their within-step iterations while preserving separate solvers~\cite{Degroote_partitioned_FSI_2010}. In this study, single-pass (weak) coupling denotes a fluid--rigid-body coupling cycle performed once per physical time step, whereas partitioned iterative (strong) coupling denotes within-step iterations of the fluid and rigid-body solvers until the kinematic residual satisfies a prescribed tolerance.

Conventional PMS implementations typically follow a single-pass sequence: the fluid-like calculation is followed by the rigid-body correction, without another fluid solve within the same time step. Although the correction enforces rigid-body motion, the pressure and surrounding-fluid response are not recomputed using the corrected velocities until a subsequent step. Consequently, the velocity obtained from the fluid-like calculation and the imposed rigid-body velocity need not be kinematically consistent at the end of the coupling pass. This mismatch can be particularly important for lightweight bodies whose motion is sensitive to the hydrodynamic response. Importantly, the single-pass character pertains to conventional PMS implementations rather than to the momentum-preserving rigid-body projection itself, which could also be incorporated into an iterative procedure.

The vulnerability of conventional partitioned coupling at low structural mass is well established in incompressible fluid--structure interaction. Causin et al.~\cite{Causin_Added-mass_partitioned_2005} showed that weakly coupled schemes can become unstable under strong added-mass effects and that strongly coupled partitioned schemes may require many subiterations as the added-mass effect becomes more pronounced. F\"orster et al.~\cite{Forster_added-mass_FSI_2007} further showed that the sequential staggered coupling analyzed in their study exhibits a limiting fluid-to-structural mass ratio for stability and that reducing the time-step size does not, in general, eliminate the associated added-mass instability. For rigid bodies, Banks et al.~\cite{Banks_FSI_AMP_2017} demonstrated that this difficulty can alternatively be addressed without subiterations through specifically designed added-mass partitioned coupling conditions. These studies show that stability depends critically on the temporal coupling formulation and cannot, in general, be ensured by reducing the time-step size alone. These observations motivate an investigation of whether improving within-step fluid--rigid-body consistency can enhance numerical stability for lightweight bodies.

Partitioned iterative coupling has also been investigated in particle-based methods. Mitsume~\cite{Mitsume_Patitioned-Iterative_FSI_MPS_FEM_2023} presented an iterative particle--finite element coupling framework. Jia et al.~\cite{Jia_iterative_ISPH-DEM_2026} developed an iterative ISPH--DEM framework for multiple arbitrarily shaped rigid bodies, including challenging configurations with narrow inter-body gaps and different fluid and solid time steps. Omura et al.~\cite{Omura_Tsuji_FD-SPH_2026} established a strongly coupled fictitious-domain smoothed particle hydrodynamics (FD-SPH) framework, extending the PMS concept to deformable structures and strong coupling, with beam elements used to demonstrate the formulation.

The approaches of Jia et al. and Omura et al. differ in how fluid--structure compatibility is enforced: Jia et al. impose velocity constraints at immersed-boundary points on solid surfaces, whereas Omura et al. employ volumetric residual feedback over fictitious-domain particles within each physical time step. The latter framework was assessed primarily for deformable beams with densities comparable to or greater than that of the fluid, leaving its stability for substantially lighter structures unresolved. Furthermore, rigid-body translation and rotation through Newton--Euler updates and volume-based hydrodynamic force and torque evaluation have not been systematically assessed within this FD-SPH framework, particularly through matched single-pass and iterative comparisons for lightweight floating bodies.

Building on the strongly coupled FD-SPH framework established by Omura et al.~\cite{Omura_Tsuji_FD-SPH_2026}, the present study develops and assesses its implementation for free-surface fluid--rigid-body interaction. The formulation combines previously established volume-based approximations of hydrodynamic force and torque with Newton--Euler updates of rigid-body translation and rotation, while feeding the kinematic mismatch between the pressure-projected fictitious-fluid and rigid-body velocities back to the fluid solver through within-step partitioned iterations. As a complementary methodological clarification, we relate PMS and single-pass FD-SPH from a fractional-step perspective. Both enforce rigid-body motion following a fluid-like update and are grounded in linear and angular momentum balance. The momentum conditions under which the PMS projection reproduces the Newton--Euler rigid-body update, including its rotational contribution, are explicitly derived in Section~\ref{subsec:pms_relation}.

Four numerical examples are considered to assess different aspects of the proposed formulation. A two-dimensional settling-disk benchmark~\cite{Glowinski_DFFD_2001} provides a baseline comparison with published numerical results. Three-dimensional buoyancy-driven cube tests assess the stability of single-pass and iterative coupling at low density ratios, including time-step sensitivity, residual convergence, hydrostatic equilibrium, and stable orientations. Wet-bed dam-break experiments involving a floating body~\cite{Wang-Dong_Exp_dam-floating-box_2022} provide measurements of free-surface evolution and rigid-body motion for validation of the two- and three-dimensional simulations. Finally, a two-dimensional wet-bed dam-break simulation with two star-shaped rigid bodies demonstrates the applicability of the formulation to multiple bodies with complex geometries and evolving inter-body flows.

The remainder of this paper is organized as follows. \Cref{sec:gov_eqs} presents the governing equations and the FD formulation. \Cref{sec:numerical_formulation} describes the partitioned iterative algorithm, acceleration strategies, and relationship to PMS. \Cref{sec:numerical_examples} presents the numerical assessments, experimental validation, and demonstration involving multiple rigid bodies with complex geometries. \Cref{sec:conclusions} summarizes the main findings.


\section{Governing equations and fictitious domain formulation for fluid--rigid-body interaction}
\label{sec:gov_eqs}

This section presents the formulation of fluid--rigid-body interaction based on the fictitious domain (FD) method, which provides the mathematical foundation for the partitioned iterative FD-SPH framework~\cite{Omura_Tsuji_FD-SPH_2026}. The governing equations and interface conditions for a conventional fluid--rigid-body interaction problem are first introduced, followed by their reformulation within the FD framework.

Consider a coupled system consisting of an incompressible Newtonian fluid and a rigid body. The physical fluid domain and rigid-body domain are denoted by $\Omega_f$ and $\Omega_s$, respectively, and their interface is defined as $\Gamma_{fs}:=\partial\Omega_f\cap\partial\Omega_s $. Unless otherwise stated, $\bm n_s$ denotes the unit normal vector on $\Gamma_{fs}$ pointing outward from the rigid-body domain $\Omega_s$ toward the fluid domain $\Omega_f$.

\begin{figure}[tb]
    \centering
    \includegraphics[width=0.95\linewidth]{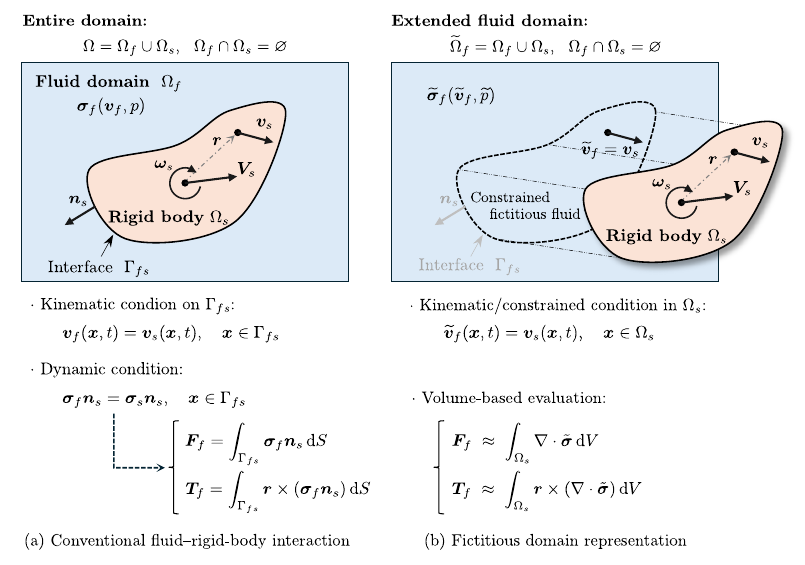}
    \caption{
    Schematic comparison of (a) conventional fluid--rigid-body interaction and (b) fictitious-domain representation, contrasting interface conditions and surface-traction evaluation with volumetric constraints and volume-based load approximation.
    }
    \label{fig:fd_domain_concept}
\end{figure}

Figure~\ref{fig:fd_domain_concept} schematically compares the conventional fluid--rigid-body interaction formulation with its fictitious-domain (FD) representation. In the conventional formulation, kinematic compatibility and hydrodynamic loads are imposed and evaluated on the fluid--rigid-body interface. In contrast, the FD formulation extends the fluid field into the rigid-body domain, where rigid-body motion is imposed as a volumetric constraint and hydrodynamic loads are approximated through volume-based evaluation.

\subsection{Governing equations and interface conditions}
\label{subsec:governing_equations}

The fluid--rigid-body interaction problem consists of the governing equations for the incompressible fluid and rigid-body motion together with the kinematic and dynamic conditions imposed at the fluid--rigid-body interface.


\paragraph{Incompressible fluid}

The motion of an incompressible Newtonian fluid occupying $\Omega_f$ is governed by the conservation of linear momentum and mass,
\begin{equation}
\label{eq:linear_mom}
\rho_f\frac{{\rm D}\bm v_f}{{\rm D}t} = \nabla\cdot\bm\sigma_f + \rho_f\bm g, \qquad \bm x\in\Omega_f ,
\end{equation}
\begin{equation}
\label{eq:cont_eqn}
\nabla\cdot\bm v_f = 0, \qquad \bm x\in\Omega_f .
\end{equation}
Here, $\bm v_f$, $\rho_f$, $\bm\sigma_f$, and $\bm g$ denote the fluid velocity, fluid density, Cauchy stress tensor, and gravitational acceleration, respectively. For an incompressible Newtonian fluid, the Cauchy stress tensor is given by
\begin{equation}
\label{eq:sigma_fluid}
\bm\sigma_f := -p\bm 1 + 2\mu_f\mathbf D(\bm v_f),
\end{equation}
where $p$ and $\mu_f$ denote the pressure and dynamic viscosity, respectively, with kinematic viscosity $\nu_f=\mu_f/\rho_f$. The symbol $\bm 1$ denotes the identity tensor of the appropriate dimension, and
\begin{equation}
\label{eq:vel_grad}
\mathbf D(\bm v_f) := \frac{1}{2} \left( \nabla\bm v_f + \left(\nabla\bm v_f\right)^{\top} \right)
\end{equation}
is the strain-rate tensor. For a fluid with constant viscosity, substitution of Eq.~\eqref{eq:sigma_fluid} into Eq.~\eqref{eq:linear_mom} gives
\begin{equation}
\label{eq:ns_eqn}
\rho_f\frac{{\rm D}\bm v_f}{{\rm D}t} = -\nabla p + \mu_f\nabla^2\bm v_f + \rho_f\bm g, \qquad \bm x\in\Omega_f .
\end{equation}
In the present study, the material time derivative ${\rm D}/{\rm D}t$ is evaluated within an Arbitrary Lagrangian--Eulerian (ALE) framework in order to account for particle motion inherent to the particle-based discretization.


\paragraph{Rigid-body dynamics}

Let $\bm X_s$, $\bm V_s$, and $\bm\omega_s$ denote the center-of-mass position, translational velocity, and angular velocity of the rigid body, respectively. Its translational and rotational motions are governed by
\begin{equation}
\label{eq:rigid_translational}
M_s\frac{{\rm d}\bm V_s}{{\rm d}t} = M_s\bm g + \bm F_f ,
\end{equation}
\begin{equation}
\label{eq:rigid_rotational}
\mathbf I_s\frac{{\rm d}\bm\omega_s}{{\rm d}t} + \bm\omega_s \times \left( \mathbf I_s\bm\omega_s \right) = \bm T_f ,
\end{equation}
where $M_s$ and $\mathbf I_s$ are the rigid-body mass and inertia tensor, respectively, while $\bm F_f$ and $\bm T_f$ denote the hydrodynamic force and torque exerted by the surrounding fluid.

For completeness, the rigid-body mass and inertia tensor are defined as
\begin{equation}
\label{eq:mass_rigid}
M_s = \int_{\Omega_s} \rho_s\,{\rm d}V ,
\end{equation}
\begin{equation}
\label{eq:inertia_rigid}
\mathbf I_s(t) = \int_{\Omega_s(t)} \rho_s \left[ (\bm r\cdot\bm r)\bm 1 - \bm r\otimes\bm r \right] {\rm d}V ,
\end{equation}
where $\rho_s$ is the rigid-body density and $\bm r := \bm x-\bm X_s$. Here, $\mathbf I_s(t)$ is expressed in the spatial coordinate system and therefore changes with the rigid-body orientation in general three-dimensional motion. 

Once $\bm V_s$ and $\bm\omega_s$ are known, the velocity at any point inside the rigid body is uniquely determined by
\begin{equation}
\label{eq:rigid_velocity_field}
\bm v_s(\bm x,t) = \bm V_s(t) + \bm\omega_s(t)\times\bm r, \qquad \bm x\in\Omega_s .
\end{equation}


\paragraph{Interface conditions}

The fluid and rigid-body motions are coupled through kinematic and dynamic conditions imposed on $\Gamma_{fs}$. The kinematic condition requires continuity of velocity,
\begin{equation}
\label{eq:kinematic_condition}
\bm v_f(\bm x,t) = \bm v_s(\bm x,t), \qquad \bm x\in\Gamma_{fs},
\end{equation}
which simultaneously enforces the no-penetration and no-slip conditions.

The dynamic condition is expressed through the traction transmitted across the interface. For a general fluid--solid system, traction continuity may be written as
\begin{equation}
\label{eq:dynamic_condition_general}
\bm\sigma_f\bm n_s = \bm\sigma_s\bm n_s, \qquad \bm x\in\Gamma_{fs},
\end{equation}
where $\bm\sigma_s$ denotes the solid stress tensor. For a perfectly rigid body, the internal stress field is not explicitly determined because the body undergoes no deformation and no constitutive stress--strain relation is solved. Instead, the interaction with the surrounding fluid is represented by the resultant hydrodynamic force and torque obtained from the fluid traction acting on the rigid-body surface,
\begin{equation}
\label{eq:fluid_force_surface}
\bm F_f = \int_{\Gamma_{fs}} \bm\sigma_f\bm n_s \,{\rm d}S ,
\end{equation}
\begin{equation}
\label{eq:fluid_torque_surface}
\bm T_f = \int_{\Gamma_{fs}} \bm r \times \left( \bm\sigma_f\bm n_s \right) {\rm d}S .
\end{equation}
Equations~\eqref{eq:fluid_force_surface} and~\eqref{eq:fluid_torque_surface} provide the conventional boundary-force representation of fluid--rigid-body interaction. Although physically straightforward, their numerical evaluation requires an explicit representation of the moving fluid--rigid-body interface together with appropriate evaluation of surface tractions and normal vectors. The FD formulation introduced below avoids this explicit boundary-force treatment by extending the fluid equations into the rigid-body domain.


\subsection{Fictitious domain formulation}
\label{subsec:FD_formulation}

In the FD formulation, the physical fluid domain is fictitiously extended into the rigid-body domain~\cite{Glowinski_LM_FD_1999,Diazgoano_FD_FEM_2003}. The extended fluid domain is defined as
\begin{equation}
\label{eq:extended_fluid_domain}
\widetilde{\Omega}_f := \Omega_f\cup\Omega_s .
\end{equation}
The rigid-body region $\Omega_s$ is therefore occupied by a fictitious fluid governed by the same constitutive relation as the surrounding physical fluid. The extended velocity, pressure, and Cauchy stress fields are denoted by $\widetilde{\bm v}_f$, $\widetilde p$, and $\widetilde{\bm\sigma}_f$, respectively, with
\begin{equation}
\label{eq:extended_stress}
\widetilde{\bm\sigma}_f := -\widetilde p\bm 1 + 2\mu_f\mathbf D(\widetilde{\bm v}_f), \qquad \bm x\in\widetilde{\Omega}_f .
\end{equation}
In the physical fluid region, the extended fields coincide with the original fluid variables,
\begin{equation}
\label{eq:extended_field_physical}
\widetilde{\bm v}_f\big|_{\Omega_f} = \bm v_f, \qquad \widetilde p\big|_{\Omega_f} = p .
\end{equation}

The fictitious fluid occupying the rigid-body domain $\Omega_s$ is required to follow the motion of the corresponding rigid body~\cite{Glowinski_LM_FD_1999}. Accordingly, the following kinematic constraint is imposed throughout the rigid-body domain:
\begin{equation}
\label{eq:fd_velocity_constraint}
\widetilde{\bm v}_f(\bm x,t) = \bm v_s(\bm x,t) = \bm V_s(t) + \bm\omega_s(t)\times\bm r, \qquad \bm x\in\Omega_s .
\end{equation}
Thus, the interface condition of Eq.~\eqref{eq:kinematic_condition} is replaced in the FD formulation by a volumetric constraint imposed over the entire rigid-body domain.

\paragraph{Volumetric constraint force}

To enforce Eq.~\eqref{eq:fd_velocity_constraint}, a volumetric constraint force is introduced in the rigid-body subdomain of the extended fluid domain. Following the FD formulation~\cite{Glowinski_LM_FD_1999,Omura_Tsuji_FD-SPH_2026}, the momentum equation over the extended domain is written as
\begin{equation}
\label{eq:extended_momentum_with_lambda}
\rho_f\frac{{\rm D}\widetilde{\bm v}_f}{{\rm D}t} = \nabla\cdot\widetilde{\bm\sigma}_f + \rho_f\bm g - \widetilde{\bm\lambda}, \qquad \bm x\in\widetilde{\Omega}_f ,
\end{equation}
together with
\begin{equation}
\label{eq:extended_continuity}
\nabla\cdot\widetilde{\bm v}_f = 0, \qquad \bm x\in\widetilde{\Omega}_f .
\end{equation}
Here, $\widetilde{\bm\lambda}$ denotes the volumetric reaction field enforcing the rigid-body kinematic constraint. With the sign convention adopted in Eq.~\eqref{eq:extended_momentum_with_lambda}, $-\widetilde{\bm\lambda}$ represents the constraint-force density acting on the fictitious fluid. At the continuum level, $\widetilde{\bm\lambda}$ is determined together with the fictitious-fluid and rigid-body motions so that the constraint $\widetilde{\bm v}_f=\bm v_s$ is satisfied throughout $\Omega_s$. Outside the rigid-body domain,
\begin{equation}
\label{eq:lambda_support}
\widetilde{\bm\lambda} = \bm 0, \qquad \bm x\in\Omega_f .
\end{equation}
In the numerical formulation presented in Section~\ref{sec:numerical_formulation}, constraint feedback is implemented through successive residual-based corrections of the velocity iterate inside the rigid-body domain. Each correction is expressed in force-density units as an iteration-wise increment, and its accumulated contribution enters the discrete fluid momentum balance. The relation between these increments and the velocity update is established algebraically in Section~\ref{subsec:corrector2}, with the viscous term evaluated using the velocity iterate specified by the numerical scheme.

Using Eq.~\eqref{eq:extended_stress}, Eq.~\eqref{eq:extended_momentum_with_lambda} may equivalently be written as
\begin{equation}
\label{eq:extended_ns_with_lambda}
\rho_f\frac{{\rm D}\widetilde{\bm v}_f}{{\rm D}t} = -\nabla\widetilde p + \mu_f\nabla^2\widetilde{\bm v}_f + \rho_f\bm g - \widetilde{\bm\lambda}, \qquad \bm x\in\widetilde{\Omega}_f .
\end{equation}
Within $\Omega_s$, the reaction field may therefore be formally expressed as
\begin{equation}
\label{eq:lambda_glowinski}
\widetilde{\bm\lambda} = -\nabla\widetilde p + \mu_f\nabla^2\widetilde{\bm v}_f + \rho_f\bm g - \rho_f\frac{{\rm D}\widetilde{\bm v}_f}{{\rm D}t}, \qquad \bm x\in\Omega_s .
\end{equation}
This reaction enforces the volumetric constraint $\widetilde{\bm v}_f=\bm v_s$ throughout $\Omega_s$.


\paragraph{Hydrodynamic force and torque}

The hydrodynamic force and torque exerted by the surrounding fluid on a rigid body are defined by the surface integrals of the fluid traction acting on the fluid--solid interface $\Gamma_{fs}$, as shown in Eqs.~\eqref{eq:fluid_force_surface} and~\eqref{eq:fluid_torque_surface}, respectively.

In the present FD formulation, the resultant hydrodynamic force and torque are approximated by volume-based evaluation, following Omura et al.~\cite{Omura_Tsuji_FD-SPH_2026}:
\begin{equation}
\label{eq:fluid_force_volume}
\bm F_f \approx \int_{\Omega_s} \nabla\cdot\widetilde{\bm\sigma}_f \,{\rm d}V,
\end{equation}
\begin{equation}
\label{eq:fluid_torque_volume}
\bm T_f \approx \int_{\Omega_s} \bm r \times \left(\nabla\cdot\widetilde{\bm\sigma}_f\right)\,{\rm d}V.
\end{equation}
These expressions motivate the discrete load evaluation rather than asserting an exact conversion of the physical surface tractions. In particular, rigid motion has zero strain rate, $\mathbf D(\bm v_s)=\bm0$, so the continuum viscous stress evaluated solely inside the rigid body cannot represent the exterior viscous traction.

The extended momentum balance relates the stress divergence to the constraint reaction through
\begin{equation}
\label{eq:extended_stress_divergence_constraint}
\nabla\cdot\widetilde{\bm\sigma}_f = \rho_f\frac{{\rm D}\widetilde{\bm v}_f}{{\rm D}t}-\rho_f\bm g+\widetilde{\bm\lambda}.
\end{equation}
Thus, the stress-divergence contribution used in the volume-based load evaluation includes both the constraint reaction and the inertial and gravitational contributions of the fictitious fluid. Since the rigid-body equations use the physical mass and inertia together with the hydrodynamic force and torque, no additional constraint-reaction load is added to those equations.

The SPH pressure-gradient and viscous operators use neighbors from both the physical- and fictitious-fluid regions, including across the fluid--rigid-body interface. The resulting force densities are integrated over the rigid-body particles to approximate the hydrodynamic force and torque (Eqs.~\eqref{eq:local_hydrodynamic_force}--\eqref{eq:fluid_torque_volume_disc}), without explicit surface integration or interface-normal evaluation.


\section{Numerical formulation of partitioned iterative FD-SPH}
\label{sec:numerical_formulation}

Building on the partitioned iterative FD-SPH framework of Omura et al.~\cite{Omura_Tsuji_FD-SPH_2026}, this section presents its rigid-body-specific formulation. The extended fluid equations are discretized using ALE-SPH and solved by a predictor--pressure-projection procedure. At each coupling iteration, the resulting hydrodynamic loads update the Newton--Euler motion; any remaining kinematic mismatch is returned to the next fluid solve until the prescribed residual criterion is met. Fixed relaxation, Aitken relaxation, and Anderson acceleration are examined.


\subsection{ALE formulation of the extended governing equations}
\label{subsec:ale_governing}

\paragraph{ALE kinematics}

The fictitious-domain governing equations presented in Section~\ref{subsec:FD_formulation} are reformulated within an Arbitrary Lagrangian--Eulerian (ALE) framework to accommodate the motion of the computational particles while allowing particle shifting in the physical-fluid domain. Introducing the transport velocity $\bm w$ of the computational particle configuration, the relative convective velocity is defined as
\begin{equation}
\label{eq:relative_velocity}
\bm c=\widetilde{\bm v}_f-\bm w .
\end{equation}
The transport velocity is defined differently in the physical-fluid and rigid-body domains as
\begin{equation}
\label{eq:ale_transport_velocity}
\bm w=
\begin{cases}
\displaystyle \widetilde{\bm v}_f+\dfrac{\delta\bm r}{\Delta t},
    & \bm x\in\Omega_f,\\[3mm]
\bm v_s, & \bm x\in\Omega_s,
\end{cases}
\end{equation}
where $\delta\bm r$ is the particle-shifting displacement prescribed for one physical time step $\Delta t$, as detailed in \ref{subsec:particle_shifting}. No independent particle shifting is applied inside the rigid-body domain; the particles in $\Omega_s$ are transported according to the rigid-body motion.

Using Eq.~\eqref{eq:relative_velocity}, the material derivative is expressed in ALE form as
\begin{equation}
\label{eq:material_derivative_ALE}
\frac{{\rm D}\widetilde{\bm v}_f}{{\rm D}t}
= \left.\frac{\partial\widetilde{\bm v}_f}{\partial t}\right|_{\mathrm{ALE}}
+ (\bm c\cdot\nabla)\widetilde{\bm v}_f.
\end{equation}
Here, the ALE time derivative follows the computational particles moving with transport velocity $\bm w$.

\paragraph{Extended-domain momentum equation}

Accordingly, the extended-domain momentum equation in Eq.~\eqref{eq:extended_momentum_with_lambda} is written in ALE form as
\begin{equation}
\label{eq:extended_momentum_ALE}
\rho_f \left[
\left.\frac{\partial\widetilde{\bm v}_f}{\partial t}\right|_{\mathrm{ALE}}
+ (\bm c\cdot\nabla)\widetilde{\bm v}_f
\right]
= -\nabla\widetilde p + \mu_f\nabla^2\widetilde{\bm v}_f
+ \rho_f\bm g - \widetilde{\bm\lambda},
\qquad \bm x\in\widetilde\Omega_f.
\end{equation}
Here, $\widetilde{\bm\lambda}$ is the volumetric constraint reaction introduced in Section~\ref{subsec:FD_formulation}, which is supported only in the rigid-body domain $\Omega_s$. The incompressibility and rigid-body kinematic constraints remain those given in Eqs.~\eqref{eq:extended_continuity} and \eqref{eq:fd_velocity_constraint}, respectively.

Because $\bm w=\bm v_s$ in $\Omega_s$ and the rigid-body constraint requires $\widetilde{\bm v}_f=\bm v_s$, Eq.~\eqref{eq:relative_velocity} gives
\begin{equation}
\label{eq:relative_velocity_solid}
\bm c=\bm0, \qquad \bm x\in\Omega_s .
\end{equation}
Thus, once the rigid-body constraint is satisfied, no relative ALE advection occurs between the fictitious-fluid velocity and the computational particle motion inside the rigid-body domain.

\paragraph{SPH particle representation}

For the numerical discretization, the extended domain is represented by SPH particles. Let
\begin{equation}
\label{eq:particle_sets}
\mathcal P_f=\left\{i\mid\bm x_i\in\Omega_f\right\}, \qquad \mathcal P_s=\left\{i\mid\bm x_i\in\Omega_s\right\},
\end{equation}
denote the sets of physical-fluid particles and particles occupying the rigid-body domain, respectively. The extended fluid equations are discretized over the unified particle set $\mathcal P=\mathcal P_f\cup\mathcal P_s$, and identical SPH differential operators are applied throughout the extended domain. The particles in $\mathcal P_s$ simultaneously serve as fictitious-fluid particles in the extended fluid calculation and as quadrature points for the rigid-body calculation, as described in Section~\ref{subsec:rigid_body_dynamics}.

To improve spatial consistency under irregular particle distributions, the second-order consistent SPH formulation, referred to as SPH(2)~\cite{Asai_SPH2_2023,Tsuji_SPH2_DPS_seepage_2025}, is adopted for the differential operators. The corresponding SPH discretizations, together with the particle-regularization and stabilization procedures, including particle shifting, are summarized in \ref{subsec:SPH_discretization}.


\subsection{Partitioned coupling procedure}
\label{subsec:fractional_step}

\paragraph{Coupling iteration}

Let $t^N$ and $t^{N+1}=t^N+\Delta t$ denote the current and next physical time levels, respectively, and let the superscript $(k)$ denote the $k$-th coupling iteration performed within the same physical time step. The velocity iterate supplied to the first fluid calculation is initialized as
\begin{equation}
\label{eq:initial_velocity_iteration}
\widetilde{\bm v}_f^{N+1(0)}=\widetilde{\bm v}_f^N .
\end{equation}
During the coupling procedure, $\widetilde{\bm v}_f^{N+1(k)}$ denotes the velocity iterate returned to the subsequent fluid solve and should therefore be distinguished from the pressure-projected fluid solution $\widetilde{\bm v}_f^{**(k)}$ and the finally accepted velocity $\widetilde{\bm v}_f^{N+1}$. The coupling index $k$ does not represent an additional physical-time advance.

\paragraph{Time-discrete extended momentum equation}

At coupling iteration $k$, the predictor and pressure projection together give the following discrete momentum balance:
\begin{align}
\label{eq:discrete_fluid_momentum}
\rho_f \frac{\widetilde{\bm v}_f^{**(k)}-\widetilde{\bm v}_f^{N+1(k-1)}}{\Delta t}
={}&-\nabla\widetilde p^{N+1(k)}+\mu_f\nabla^2\widetilde{\bm v}_f^{N+1(k-1)}
+\rho_f\bm g-\rho_f(\bm c^N\cdot\nabla)\widetilde{\bm v}_f^N,
\qquad \bm x\in\widetilde\Omega_f .
\end{align}
Equivalently, expressing the temporal increment relative to the accepted velocity at $t^N$ gives
\begin{align}
\label{eq:discrete_extended_momentum}
\rho_f \frac{\widetilde{\bm v}_f^{**(k)}-\widetilde{\bm v}_f^N}{\Delta t}
={}&-\nabla\widetilde p^{N+1(k)}+\mu_f\nabla^2\widetilde{\bm v}_f^{N+1(k-1)}
+\rho_f\bm g-\rho_f(\bm c^N\cdot\nabla)\widetilde{\bm v}_f^N
+\rho_f\frac{\widetilde{\bm v}_f^{N+1(k-1)}-\widetilde{\bm v}_f^N}{\Delta t}.
\end{align}
The last term in Eq.~\eqref{eq:discrete_extended_momentum} represents the constraint feedback accumulated through the preceding coupling iterations. It vanishes in the physical-fluid domain because the velocity iterate is reset to $\widetilde{\bm v}_f^N$ there, whereas inside the rigid-body domain it contains the residual-based corrections returned from the preceding fluid--rigid-body coupling cycles. As shown later in Section~\ref{subsec:corrector2}, this term may be written as
\begin{equation}
\rho_f \frac{ \widetilde{\bm v}_f^{N+1(k-1)} - \widetilde{\bm v}_f^N }{\Delta t} = - \sum_{j=1}^{k-1} \Delta\widetilde{\bm\lambda}^{N+1(j)}, \qquad \bm x\in\Omega_s .
\end{equation}
Hence, the final term in Eq.~\eqref{eq:discrete_extended_momentum} can be interpreted as the discrete counterpart of the constraint-force contribution $-\widetilde{\bm\lambda}$ in the continuous extended-domain momentum equation.

Upon convergence, the pressure-projected fictitious-fluid velocity satisfies the rigid-body kinematic constraint within the prescribed tolerance. In the discrete momentum balance of Eq.~\eqref{eq:discrete_extended_momentum}, the accumulated constraint feedback appears explicitly in the final term and also contributes to the viscous term through the corrected velocity iterate, as follows from Eq.~\eqref{eq:cumulative_constraint_correction}. This viscous contribution is also included in the hydrodynamic force-density evaluation in Eq.~\eqref{eq:local_hydrodynamic_force} as part of the adopted time-discrete coupling scheme.

\paragraph{Fluid--rigid-body coupling cycle}

The fluid problem in Eq.~\eqref{eq:discrete_fluid_momentum} is evaluated using a two-stage fractional-step procedure. Starting from $\widetilde{\bm v}_f^{N+1(k-1)}$, the predictor gives $\widetilde{\bm v}_f^{*(k)}$, after which Corrector-1 applies the pressure projection and gives $\widetilde{\bm v}_f^{**(k)}$. The resulting pressure and viscous stress fields are used to determine the rigid-body motion, and the remaining kinematic mismatch is returned to the subsequent fluid solve when another coupling iteration is required:
\begin{equation}
\label{eq:partitioned_coupling_cycle}
\begin{aligned}
\widetilde{\bm v}_f^{N+1(k-1)}&\xrightarrow{\rm Predictor}\widetilde{\bm v}_f^{*(k)}
\xrightarrow{\rm Corrector\text{-}1}\widetilde{\bm v}_f^{**(k)}\xrightarrow{\rm rigid\ body}\bm v_s^{N+1(k)}
\xrightarrow{\rm residual\ correction}\widetilde{\bm v}_f^{N+1(k)}.
\end{aligned}
\end{equation}
The final operation in Eq.~\eqref{eq:partitioned_coupling_cycle} is not a third fractional-step correction applied directly to $\widetilde{\bm v}_f^{**(k)}$. Instead, the kinematic mismatch is used to update the preceding velocity iterate $\widetilde{\bm v}_f^{N+1(k-1)}$, thereby defining the velocity supplied to the subsequent fluid calculation. This feedback operation is omitted when the current coupling step is accepted, including the prescribed single-pass case $K_{\max}=1$; the accepted rigid-body velocity in the solid domain is then imposed according to Eq.~\eqref{eq:velocity_acceptance}, as distinguished in Section~\ref{subsec:pms_relation}.


\subsection{Predictor step}
\label{subsec:predictor}

At coupling iteration $k$, the predictor velocity is evaluated from advection, viscous diffusion, and gravity as
\begin{equation}
\label{eq:predictor_step}
\frac{ \widetilde{\bm v}_f^{*(k)} - \widetilde{\bm v}_f^{N+1(k-1)} }{\Delta t} = \frac{\mu_f}{\rho_f} \nabla^2\widetilde{\bm v}_f^{N+1(k-1)} + \bm g - (\bm c^N\cdot\nabla) \widetilde{\bm v}_f^N, \qquad \bm x\in\widetilde\Omega_f .
\end{equation}
The ALE advection term is evaluated explicitly using the accepted velocity $\widetilde{\bm v}_f^N$ and the relative convective velocity $\bm c^N$ and therefore remains fixed throughout the coupling iterations within the current physical time step. In contrast, the velocity iterate $\widetilde{\bm v}_f^{N+1(k-1)}$ enters both the temporal increment and the viscous contribution. Consequently, the constraint correction obtained from one coupling iteration directly modifies the subsequent fluid predictor and, through the pressure projection, the subsequent pressure field.

For the first coupling iteration, Eq.~\eqref{eq:initial_velocity_iteration} gives $\widetilde{\bm v}_f^{N+1(0)}=\widetilde{\bm v}_f^N$. For subsequent iterations, the physical-fluid velocity is reset to the accepted state at $t^N$, whereas the residual-based correction is retained inside the rigid-body domain, as detailed in Section~\ref{subsec:corrector2}.


\subsection{Corrector-1: Incompressibility projection}
\label{subsec:corrector1}

Corrector-1 applies a pressure projection to the predictor velocity to reduce its divergence,
\begin{equation}
\label{eq:pressure_split}
\rho_f \frac{\widetilde{\bm v}_f^{**(k)}-\widetilde{\bm v}_f^{*(k)}}{\Delta t}
=
-\nabla\widetilde p^{N+1(k)} .
\end{equation}
Imposing $\nabla\cdot\widetilde{\bm v}_f^{**(k)}=0$ and taking the divergence of Eq.~\eqref{eq:pressure_split} gives the pressure Poisson equation
\begin{equation}
\label{eq:pressure_poisson}
\nabla^2\widetilde p^{N+1(k)}
=
\frac{\rho_f}{\Delta t}
\nabla\cdot\widetilde{\bm v}_f^{*(k)},
\qquad \bm x\in\widetilde\Omega_f .
\end{equation}
The free-surface particle set $\mathcal P_{\mathrm{FS}}$ is identified from the unified particle set $\mathcal P=\mathcal P_f\cup\mathcal P_s$ using the multistage detection procedure described in \ref{subsec:sph2_formulation}. Thus, atmosphere-exposed rigid-body particles are also treated as free-surface particles in the extended-fluid calculation. The atmospheric-pressure Dirichlet condition is imposed by assigning zero gauge pressure to these particles:
\begin{equation}
\label{eq:free_surface_pressure_bc}
\widetilde p_i^{N+1(k)}=0,
\qquad i\in\mathcal P_{\mathrm{FS}}.
\end{equation}
The pressure field $\widetilde p^{N+1(k)}$ is obtained at each coupling iteration by solving the linear system resulting from the SPH discretization of Eq.~\eqref{eq:pressure_poisson}, subject to the prescribed pressure boundary conditions, using the BiCGSTAB method. The BiCGSTAB solve is terminated when the absolute Euclidean norm of the algebraic pressure-Poisson residual falls below $10^{-6}$. This linear-solver tolerance is distinct from the relative kinematic-residual criterion used for the fluid--rigid-body coupling iteration in Section~\ref{subsec:partitioned_iteration}.

Once $\widetilde p^{N+1(k)}$ has been obtained, the pressure-projected extended-fluid velocity is evaluated as
\begin{equation}
\label{eq:pressure_projection}
\widetilde{\bm v}_f^{**(k)}
=
\widetilde{\bm v}_f^{*(k)}
-
\frac{\Delta t}{\rho_f}
\nabla\widetilde p^{N+1(k)},
\qquad \bm x\in\widetilde\Omega_f .
\end{equation}
The pressure projection aims to enforce incompressibility over the extended domain. Its accuracy depends on the pressure-solver tolerance and the consistency between the discrete divergence, gradient, and Laplacian operators.

In conventional projection methods that impose boundary conditions directly on the moving rigid-body surface, a pressure Neumann condition can be obtained from the wall-normal component of the fluid momentum equation by requiring compatibility between the fluid and rigid-body accelerations~\cite{Banks_AMP_RB_2017}. For the constant-viscosity Newtonian fluid considered here and the outward rigid-body normal $\bm n_s$ defined in Section~\ref{subsec:governing_equations}, this continuous compatibility condition is
\begin{equation}
\left.\frac{\partial p}{\partial n_s}\right|_{\Gamma_{fs}}
=
\bm n_s\cdot
\left[
\mu_f\nabla^2\bm v_f
+
\rho_f(\bm g-\bm a_\Gamma)
\right],
\end{equation}
where the acceleration at a point on the rigid-body surface is
\begin{equation}
\bm a_\Gamma
=
\frac{\mathrm d\bm V_s}{\mathrm dt}
+
\frac{\mathrm d\bm\omega_s}{\mathrm dt}\times\bm r
+
\bm\omega_s\times(\bm\omega_s\times\bm r).
\end{equation}
The terms in $\bm a_\Gamma$ describe translational, angular, and centripetal acceleration. The expression above follows from the continuous momentum balance; a time-discrete projection method may impose a related condition based on the predicted and prescribed wall-normal velocities, whose precise form depends on the splitting scheme. Banks et al.~\cite{Banks_AMP_RB_2017} used this acceleration-compatibility condition as part of an added-mass partitioned scheme. Consistent pressure boundary conditions for projection-based SPH have also been examined by Hosseini and Feng~\cite{Hosseini_Feng_PressureBC_SPH_2011} for stationary solid walls and open boundaries.

Difficulties can arise when distinct walls or sharp corners are present within a particle's interaction neighborhood, because multiple wall-normal conditions must then be represented in the discrete pressure solve. Okano et al.~\cite{OKANO2027119471} discussed this issue for conventional projection-based SPH and developed a projection-free, VMS-stabilized monolithic velocity--pressure formulation as an alternative. The present FD-SPH method instead retains the pressure projection but extends the fluid equations throughout $\widetilde\Omega_f=\Omega_f\cup\Omega_s$, so that no separate pressure Neumann condition or interface-normal evaluation is imposed on moving rigid-body boundaries. The exterior stationary walls are treated by the fixed-wall ghost-particle procedure described by Morikawa et al.~\cite{Morikawa_ALE-ISPH_DPS_2023}. This avoids explicit moving-interface pressure conditions but does not eliminate errors associated with under-resolved narrow gaps, particle support, or discrete-operator consistency.

The pressure projection targets incompressibility over the extended fluid domain but does not itself impose rigid-body kinematics. The resulting mismatch between the pressure-projected fictitious-fluid velocity and the Newton--Euler rigid-body velocity is quantified by the kinematic residual in Eq.~\eqref{eq:coupling_residual}. In iterative coupling, this residual is used to update the subsequent fluid solution, as described in Section~\ref{subsec:corrector2}.


\subsection{Hydrodynamic force evaluation and rigid-body update}
\label{subsec:rigid_body_dynamics}

\paragraph{Particle representation of the rigid body}

The rigid-body domain $\Omega_s$ is represented by the particle set $\mathcal P_s$ introduced in Eq.~\eqref{eq:particle_sets}. The particles in $\mathcal P_s$ play two distinct roles. In the extended fluid calculation, they serve as fictitious-fluid particles governed by the fluid density $\rho_f$. In the rigid-body equations of motion, the same particle locations are used as quadrature points representing the physical solid mass with density $\rho_s$.

The total rigid-body mass is evaluated as
\begin{equation}
\label{eq:mass_rigid_particle}
M_s = \int_{\Omega_s}\rho_s\,{\rm d}V \approx \sum_{i\in\mathcal P_s}\rho_s V_i = \sum_{i\in\mathcal P_s}m_{s,i}, \qquad m_{s,i}=\rho_sV_i .
\end{equation}
Because the rigid-body configuration is kept fixed throughout the coupling iterations within a physical time step, the inertia tensor is evaluated from the configuration at time $t^N$ and held fixed during the iteration,
\begin{equation}
\label{eq:inertia_rigid_particle}
\mathbf I_s^N = \int_{\Omega_s^N} \rho_s \left[ (\bm r\cdot\bm r)\bm 1-\bm r\otimes\bm r \right] {\rm d}V \approx \sum_{i\in\mathcal P_s} m_{s,i} \left[ (\bm r_i^N\cdot\bm r_i^N)\bm 1-\bm r_i^N\otimes\bm r_i^N \right],
\end{equation}
where $\bm r_i^N=\bm x_i^N-\bm X_s^N$. 
The center of mass and inertia tensor are both evaluated using the same discrete solid-particle mass distribution.
In the subsequent discrete formulation, $M_s$ and $\mathbf I_s^N$ denote the particle-quadrature values given by the rightmost expressions above.

\paragraph{Hydrodynamic force and torque}

Following the fictitious-domain formulation introduced in Section~\ref{subsec:FD_formulation}, the local hydrodynamic force density is evaluated as a discrete approximation of the divergence of the extended fictitious-fluid stress. Consistently with the predictor step, the force density at coupling iteration $k$ is evaluated as
\begin{equation}
\label{eq:local_hydrodynamic_force}
\bm f_i^{N+1(k)} = - \left( \nabla\widetilde p^{N+1(k)} \right)_i + \mu_f \left( \nabla^2\widetilde{\bm v}_f^{N+1(k-1)} \right)_i, \qquad i\in\mathcal P_s .
\end{equation}
The resultant hydrodynamic force is approximated by
\begin{equation}
\label{eq:fluid_force_volume_disc}
\bm F_f^{N+1(k)} \approx \int_{\Omega_s}\bm f^{N+1(k)}\,{\rm d}V \approx \sum_{i\in\mathcal P_s}\bm f_i^{N+1(k)}V_i ,
\end{equation}
and the hydrodynamic torque about the center of mass is evaluated as
\begin{equation}
\label{eq:fluid_torque_volume_disc}
\bm T_f^{N+1(k)} \approx \int_{\Omega_s}\bm r\times\bm f^{N+1(k)}\,{\rm d}V \approx \sum_{i\in\mathcal P_s}\bm r_i^N\times\bm f_i^{N+1(k)}V_i .
\end{equation}
The quantities in Eqs.~\eqref{eq:fluid_force_volume_disc} and \eqref{eq:fluid_torque_volume_disc} provide the discrete volume-based approximations of the physical hydrodynamic force and torque used in the Newton--Euler equations. They are evaluated from the pressure and viscous stress fields and should be distinguished from the residual-based constraint correction introduced in Section~\ref{subsec:corrector2}.

\paragraph{Rigid-body motion}

Using the hydrodynamic force and torque obtained above, the translational velocity is evaluated as
\begin{equation}
\label{eq:rigid_translational_update}
\bm V_s^{N+1(k)} = \bm V_s^N + \Delta t \left( \bm g+\frac{\bm F_f^{N+1(k)}}{M_s} \right),
\end{equation}
and the angular velocity is evaluated from the discrete Euler equation as
\begin{equation}
\label{eq:rigid_rotational_update}
\bm\omega_s^{N+1(k)} = \bm\omega_s^N + \Delta t \left( \mathbf I_s^N \right)^{-1} \left[ \bm T_f^{N+1(k)} - \bm\omega_s^N \times \left( \mathbf I_s^N\bm\omega_s^N \right) \right].
\end{equation}
The rigid-body translational and angular velocities are recomputed from the accepted state at time $t^N$ during each coupling iteration, while the rigid-body configuration and inertia tensor are kept fixed. The corresponding rigid-body velocity field is
\begin{equation}
\label{eq:rigid_velocity_field_disc}
\bm v_{s,i}^{N+1(k)} = \bm V_s^{N+1(k)} + \bm\omega_s^{N+1(k)}\times\bm r_i^N, \qquad i\in\mathcal P_s .
\end{equation}


\subsection{Residual-based rigid-body constraint correction}
\label{subsec:corrector2}

\paragraph{Kinematic residual}

After the fluid and rigid-body subproblems have been solved at coupling iteration $k$, the kinematic mismatch inside the rigid-body domain is defined as
\begin{equation}
\label{eq:coupling_residual}
\bm R^{(k)} = \widetilde{\bm v}_f^{**(k)} - \bm v_s^{N+1(k)}, \qquad \bm x\in\Omega_s .
\end{equation}
The condition $\bm R^{(k)}=\bm 0$ represents kinematic compatibility between the pressure-projected fictitious-fluid velocity and the rigid-body velocity. In the zero-residual limit, the two velocity fields coincide inside $\Omega_s$, and the rigid-body kinematic constraint is satisfied by the pressure-projected extended-fluid solution.

The residual in Eq.~\eqref{eq:coupling_residual} is evaluated after the pressure projection and rigid-body update. It therefore measures the remaining inconsistency between the fluid and rigid-body subproblems at the current coupling iteration. This residual is used both for the convergence assessment and, if a further coupling iteration is required, for constructing the correction returned to the subsequent fluid solve.

\paragraph{Generalized coupling correction}

To separate the physical role of the fictitious-domain constraint feedback from the particular convergence-acceleration technique used to determine it, a generalized coupling correction $\bm C^{(k)}$ is introduced. The velocity iterate supplied to the subsequent fluid solve is written as
\begin{equation}
\label{eq:coupling_velocity_update}
\widetilde{\bm v}_f^{N+1(k)} =
\begin{cases}
\widetilde{\bm v}_f^N, & \bm x\in\Omega_f,\\[2mm]
\displaystyle \widetilde{\bm v}_f^{N+1(k-1)} - \bm C^{(k)}, & \bm x\in\Omega_s .
\end{cases}
\end{equation}
Thus, the physical-fluid part of the velocity iterate is reset to the accepted state at $t^N$, whereas the fictitious-fluid velocity inside the rigid-body domain retains the coupling corrections accumulated during the within-step iterations.

The correction $\bm C^{(k)}$ is an algorithmic quantity determined from the kinematic residual and the selected convergence-acceleration strategy. For a scalar relaxation scheme,
\begin{equation}
\label{eq:scalar_coupling_correction}
\bm C^{(k)} = \omega^{(k)}\bm R^{(k)},
\end{equation}
which recovers the conventional fixed or dynamically relaxed update. More generally, $\bm C^{(k)}$ may contain information from several preceding coupling iterates and residuals, as in the Anderson acceleration described in Section~\ref{subsec:partitioned_iteration}.

Equation~\eqref{eq:coupling_velocity_update} emphasizes that the coupling correction is applied to the preceding velocity iterate $\widetilde{\bm v}_f^{N+1(k-1)}$, rather than directly replacing the pressure-projected velocity $\widetilde{\bm v}_f^{**(k)}$. The corrected iterate subsequently enters the temporal and viscous terms of the predictor in Eq.~\eqref{eq:predictor_step}. Consequently, the mismatch obtained in one coupling cycle modifies the subsequent fluid momentum balance, pressure field, hydrodynamic loads, and rigid-body motion.

\paragraph{Equivalent constraint-force representation}

The velocity correction may be expressed in force-density units by defining the iteration-wise feedback increment
\begin{equation}
\label{eq:delta_lambda}
\Delta\widetilde{\bm\lambda}^{N+1(k)} =
\begin{cases}
\bm0, & \bm x\in\Omega_f,\\[2mm]
\displaystyle \frac{\rho_f}{\Delta t} \bm C^{(k)}, & \bm x\in\Omega_s .
\end{cases}
\end{equation}
Combining Eqs.~\eqref{eq:coupling_velocity_update} and \eqref{eq:delta_lambda} gives, inside the rigid-body domain,
\begin{equation}
\label{eq:coupling_velocity_update_lambda}
\rho_f \frac{ \widetilde{\bm v}_f^{N+1(k)} - \widetilde{\bm v}_f^{N+1(k-1)} }{ \Delta t } = - \Delta\widetilde{\bm\lambda}^{N+1(k)}, \qquad \bm x\in\Omega_s .
\end{equation}
Starting from the initialization $\widetilde{\bm v}_f^{N+1(0)}=\widetilde{\bm v}_f^N$,
the cumulative correction after $k$ coupling updates therefore satisfies
\begin{equation}
\label{eq:cumulative_constraint_correction}
\sum_{j=1}^{k} \Delta\widetilde{\bm\lambda}^{N+1(j)} = \frac{\rho_f}{\Delta t} \left( \widetilde{\bm v}_f^N - \widetilde{\bm v}_f^{N+1(k)} \right), \qquad \bm x\in\Omega_s .
\end{equation}
Thus, the accumulated feedback is represented by the recursively updated fictitious-fluid velocity, without independently storing the increments. Equation~\eqref{eq:cumulative_constraint_correction} is an algebraic identity for the adopted coupling updates and explains the feedback term in Eq.~\eqref{eq:discrete_extended_momentum}. 

The increment $\Delta\widetilde{\bm\lambda}^{N+1(k)}$ in Eq.~\eqref{eq:delta_lambda} is an algorithmic representation of the correction returned between the partitioned subproblems. It should therefore be distinguished from the physical hydrodynamic force density used to evaluate the resultant force and torque in Section~\ref{subsec:rigid_body_dynamics}.

At coupling iteration $k$, the fluid momentum balance includes the accumulated feedback $\sum_{j=1}^{k-1}\Delta\widetilde{\bm\lambda}^{N+1(j)}$, with the empty sum taken as zero for $k=1$. Once the residual passes the convergence test, the current coupled solution is accepted without constructing another feedback increment. The previously accumulated feedback remains included in the fluid solution.


\subsection{Partitioned iteration and convergence acceleration}
\label{subsec:partitioned_iteration}

\paragraph{Within-step fixed-point iteration}

For $K_{\max}>1$, the sequence in Eq.~\eqref{eq:partitioned_coupling_cycle} is repeated within the same physical time step. The corrected velocity obtained from Eq.~\eqref{eq:coupling_velocity_update} is supplied to the subsequent predictor without advancing the particle configuration or rigid-body position. Because the fictitious-domain correction enters the temporal increment and viscous contribution of Eq.~\eqref{eq:predictor_step}, it directly modifies the subsequent extended-fluid solution. The resulting change in the pressure projection modifies the hydrodynamic force and torque and, consequently, the rigid-body velocity obtained in the next coupling cycle.

This fixed-point iteration acts on the fictitious-domain velocity degrees of freedom, assembled into the following coupling vectors.

\paragraph{Particle-space coupling vectors}

Let $N_s=|\mathcal P_s|$ be the number of fictitious-domain particles, and let $d\in\{2,3\}$ be the number of velocity components per particle. Arrange $\mathcal P_s=\{i_1,\ldots,i_{N_s}\}$ in a fixed order throughout each physical time step. The coupling vector entering the $k$-th fluid solve is
\begin{equation}
\label{eq:coupling_vector}
\bm u^{(k)}
=\operatorname{col}\left(
\widetilde{\bm v}_{f,i_1}^{N+1(k-1)},
\ldots,
\widetilde{\bm v}_{f,i_{N_s}}^{N+1(k-1)}
\right)
=
\begin{bmatrix}
\widetilde{\bm v}_{f,i_1}^{N+1(k-1)}\\
\vdots\\
\widetilde{\bm v}_{f,i_{N_s}}^{N+1(k-1)}
\end{bmatrix}
\in\mathbb R^{dN_s}.
\end{equation}
Similarly, the particle-space residual is defined by stacking the kinematic residuals in Eq.~\eqref{eq:coupling_residual},
\begin{equation}
\label{eq:coupling_residual_vector}
\bm r_{\mathrm{cpl}}^{(k)} =\operatorname{col}\left(\bm R_{i_1}^{(k)},\ldots,\bm R_{i_{N_s}}^{(k)}\right).
\end{equation}
The corresponding stacked correction vector is
\begin{equation}
\label{eq:coupling_correction_vector}
\bm z^{(k)} =\operatorname{col}\left(\bm C_{i_1}^{(k)},\ldots,\bm C_{i_{N_s}}^{(k)}\right).
\end{equation}
With each entry corresponding to the particle correction $\bm C_i^{(k)}$, the fictitious-domain update in Eq.~\eqref{eq:coupling_velocity_update} becomes
\begin{equation}
\label{eq:coupling_vector_update}
\bm u^{(k+1)} =\bm u^{(k)} -\bm z^{(k)}.
\end{equation}
The discrete inner product between two particle-space vectors is defined as
\begin{equation}
\label{eq:aitken_inner_product}
\left\langle\bm a,\bm b\right\rangle_s
=\sum_{i\in\mathcal P_s}\bm a_i\cdot\bm b_i,
\end{equation}
which is the Euclidean inner product of the stacked vectors.

The coupling vectors contain velocity degrees of freedom throughout the fictitious rigid-body domain. Established acceleration algorithms are applied to this volumetric FD-SPH iteration through the common correction vector. 
For multiple rigid bodies, the coupling variables and kinematic residuals of all fictitious-domain particles are assembled into a single global vector. The relative-residual convergence criterion and Anderson acceleration are applied to this global vector, while the Newton--Euler equations are evaluated separately for each rigid body.

\paragraph{Static relaxation}

The simplest update employs a prescribed constant relaxation parameter $\omega$,
\begin{equation}
\label{eq:static_relaxation}
\bm z^{(k)} = \omega\bm r_{\mathrm{cpl}}^{(k)},
\end{equation}
which gives $\bm u^{(k+1)} = \bm u^{(k)} - \omega\bm r_{\mathrm{cpl}}^{(k)} $. The choice $\omega=1$ corresponds to the unrelaxed fixed-point iteration, whereas $0<\omega<1$ provides constant under-relaxation. Static relaxation is used as a reference strategy for assessing the convergence behavior of the dynamically accelerated schemes.

\paragraph{Aitken dynamic relaxation}

Aitken's $\Delta^2$ method determines a scalar relaxation parameter from the change in the coupling residual. The update retains the form
\begin{equation}
\label{eq:aitken_update}
\bm z^{(k)} = \omega^{(k)}\bm r_{\mathrm{cpl}}^{(k)} .
\end{equation}
At the first coupling iteration, $\omega^{(1)}=\omega_0=0.2$. For $k\ge2$, define $\Delta\bm r_{\mathrm{cpl}}^{(k)} = \bm r_{\mathrm{cpl}}^{(k)} -\bm r_{\mathrm{cpl}}^{(k-1)}.$ The relaxation parameter is then evaluated as
\begin{equation}
\label{eq:aitken}
\omega^{(k)} = - \omega^{(k-1)} \frac{ \left\langle \bm r_{\mathrm{cpl}}^{(k-1)}, \Delta\bm r_{\mathrm{cpl}}^{(k)} \right\rangle_s }{ \left\langle \Delta\bm r_{\mathrm{cpl}}^{(k)}, \Delta\bm r_{\mathrm{cpl}}^{(k)} \right\rangle_s }.
\end{equation}
If $\left\langle\Delta\bm r_{\mathrm{cpl}}^{(k)},\Delta\bm r_{\mathrm{cpl}}^{(k)} \right\rangle_s=0$, the relaxation parameter from the preceding iteration is retained. To prevent excessively small or large corrections, the value is bounded according to
\begin{equation}
\label{eq:omega_limit}
0.05 \le \omega^{(k)} \le 1.2 .
\end{equation}
Such bounds are a numerical safeguard also used in partitioned FSI schemes~\cite{Degroote_partitioned_FSI_2010}; the present limits are implementation choices.

\paragraph{Anderson acceleration}

To incorporate information from several preceding coupling iterations, a limited-memory Anderson acceleration is also employed. Anderson acceleration is a general fixed-point acceleration technique and is closely related to multisecant quasi-Newton methods~\cite{Anderson_1965,Walker_Ni_2011,Fang_Saad_2009}. In the present formulation, it is applied directly to the particle-space fictitious-domain velocity and residual vectors defined in Eqs.~\eqref{eq:coupling_vector} and \eqref{eq:coupling_residual_vector}.

For $j\ge2$, define the changes in the coupling iterate and residual as
\begin{align}
\label{eq:anderson_dq}
\Delta\bm u^{(j)} &= \bm u^{(j)} -\bm u^{(j-1)}, \\
\label{eq:anderson_dr}
\Delta\bm r_{\mathrm{cpl}}^{(j)} &= \bm r_{\mathrm{cpl}}^{(j)} -\bm r_{\mathrm{cpl}}^{(j-1)}.
\end{align}
Let $m_k=\min(m,k-1)$ denote the number of available history vectors, where $m$ is the prescribed maximum history size. The corresponding history matrices are
\begin{align}
\label{eq:anderson_DQ}
\Delta\mathbf U^{(k)} &= \left[ \Delta\bm u^{(k-m_k+1)}, \ldots, \Delta\bm u^{(k)} \right], \\
\label{eq:anderson_DR}
\Delta\mathbf R_{\mathrm{cpl}}^{(k)} &= \left[ \Delta\bm r_{\mathrm{cpl}}^{(k-m_k+1)}, \ldots, \Delta\bm r_{\mathrm{cpl}}^{(k)} \right].
\end{align}
Accordingly, $\bm u^{(k)},\bm r_{\mathrm{cpl}}^{(k)},\bm z^{(k)} \in\mathbb R^{dN_s}$ and $\Delta\mathbf U^{(k)},\Delta\mathbf R_{\mathrm{cpl}}^{(k)} \in\mathbb R^{dN_s\times m_k}$; each history difference is a column of the corresponding matrix.

In the ideal Type-II Anderson formulation, the coefficient vector $\bm\gamma^{(k)} $ is defined by
\begin{equation}
\label{eq:anderson_least_squares}
\bm\gamma^{(k)}  = \arg\min_{\bm\gamma\in\mathbb R^{m_k}} \left\| \bm r_{\mathrm{cpl}}^{(k)} - \Delta\mathbf R_{\mathrm{cpl}}^{(k)} \bm\gamma \right\|_2 .
\end{equation}
With the residual convention adopted here, $\bm r_{\mathrm{cpl}}^{(k)} = \bm u^{(k)} -\mathcal G(\bm u^{(k)}),$ where $\mathcal G(\bm u^{(k)})=\bm u^{(k)} -\bm r_{\mathrm{cpl}}^{(k)} $ denotes the unrelaxed residual-correction map defined by one fluid and rigid-body coupling cycle, the Type-II Anderson update is
\begin{equation}
\label{eq:anderson_update}
\bm u^{(k+1)} = \bm u^{(k)} - \beta\bm r_{\mathrm{cpl}}^{(k)} - \left( \Delta\mathbf U^{(k)} - \beta\Delta\mathbf R_{\mathrm{cpl}}^{(k)} \right) \bm\gamma^{(k)}.
\end{equation}
Accordingly, the generalized coupling correction in Eq.~\eqref{eq:coupling_vector_update} is
\begin{equation}
\label{eq:anderson_correction}
\bm z^{(k)} = \beta\bm r_{\mathrm{cpl}}^{(k)} + \left( \Delta\mathbf U^{(k)} - \beta\Delta\mathbf R_{\mathrm{cpl}}^{(k)} \right) \bm\gamma^{(k)}.
\end{equation}
Thus, unlike static and Aitken relaxation, Anderson acceleration modifies both the magnitude and direction of the coupling correction by using differences between preceding coupling iterates and residuals.

In the present calculations, the history size is limited to $m=3,$ and the mixing parameter is $\beta=1$. At the first coupling iteration, where no history information is available, the update reduces to fixed under-relaxation,
\begin{equation}
\label{eq:anderson_initial_update}
\bm u^{(2)} = \bm u^{(1)} - \omega_0\bm r_{\mathrm{cpl}}^{(1)}, \qquad \omega_0=0.2 .
\end{equation}
The same under-relaxed update is used as a fallback if the small Anderson least-squares system cannot be solved reliably. The Anderson history is constructed independently within each physical time step and is not carried over to the subsequent time step.

\paragraph{Solution of the Anderson least-squares problem}

The Anderson coefficients are obtained using the Gram matrix
\begin{equation}
\label{eq:anderson_gram}
\mathbf G^{(k)} = \left(\Delta\mathbf R_{\mathrm{cpl}}^{(k)}\right)^{\top}\Delta\mathbf R_{\mathrm{cpl}}^{(k)} 
\end{equation}
and right-hand side
\begin{equation}
\label{eq:anderson_rhs}
\bm b^{(k)} = \left(\Delta\mathbf R_{\mathrm{cpl}}^{(k)}\right)^{\top}\bm r_{\mathrm{cpl}}^{(k)} .
\end{equation}
This Gram-matrix approach follows a standard implementation of Anderson acceleration~\cite{Peng_Anderson_2018}.

The regularized normal equations are
\begin{equation}
\label{eq:anderson_normal_equation}
\left(\mathbf G^{(k)} +\eta_k\bm 1\right)\bm\gamma^{(k)} =\bm b^{(k)},
\end{equation}
where $\bm 1$ is the identity matrix of order $m_k$.
The regularization parameter is defined as
\begin{equation}
\label{eq:anderson_regularization}
\eta_k
=
10^{-10}
\max_{1\le j\le m_k}
\left|(\mathbf G^{(k)})_{jj}\right|.
\end{equation}
The regularization improves robustness for nearly dependent residual differences.

With $m=3$, the system is at most $3\times3$. If it cannot be solved reliably, the update falls back to under-relaxation with $\omega_0=0.2$.

\paragraph{Convergence criterion}
Convergence of the partitioned iteration is monitored using the discrete $\ell_2$ norm of the kinematic residual,
\begin{equation}
\label{eq:residual_norm}
\mathcal{R}^{(k)} = \left( \sum_{i\in\mathcal P_s} \left| \bm R_i^{(k)} \right|^2 \right)^{1/2}.
\end{equation}
The absolute residual norm $\mathcal{R}^{(k)}$ has velocity units.
For an iterative calculation, the relative residual is defined with respect to the residual obtained at the first coupling iteration within the same physical time step as
\begin{equation}
\label{eq:relative_residual}
\widehat{\mathcal{R}}^{(k)} = \frac{\mathcal{R}^{(k)}}{\mathcal{R}^{(1)}},
\qquad
\mathcal{R}^{(1)}>\mathcal{R}_{\mathrm{floor}}.
\end{equation}
Convergence is assessed exclusively using the relative residual,
\begin{equation}
\label{eq:convergence}
\widehat{\mathcal{R}}^{(k)} \le \varepsilon_{\rm rel},
\qquad
\varepsilon_{\rm rel}=10^{-3}.
\end{equation}
The numerical guard $\mathcal{R}_{\mathrm{floor}}=10^{-10}~\mathrm{cm/s}$ prevents division by a nearly zero initial residual. If $\mathcal{R}^{(1)}\le\mathcal{R}_{\mathrm{floor}}$, the coupled state is accepted after the first cycle without evaluating the relative residual; this guard is not an independent absolute convergence criterion.

After each fluid--rigid-body solve, Eq.~\eqref{eq:convergence} is checked before applying any further correction. Once satisfied, the current solution is accepted without an additional acceleration update. Prescribed single-pass (weak) coupling ($K_{\max}=1$) terminates after one cycle regardless of this test. For $K_{\max}>1$, $K_{\max}$ is only an iteration limit: if the residual has not converged and $k<K_{\max}$, the correction $\bm C^{(k)}$ is returned through Eq.~\eqref{eq:coupling_velocity_update}; if $k=K_{\max}$, the state is retained without further correction and explicitly classified as nonconverged.


\subsection{Acceptance of the coupled solution and particle update}
\label{subsec:solution_acceptance}

\paragraph{Accepted velocity field}

When the coupling iteration terminates, the accepted extended-fluid velocity is assembled from the pressure-projected velocity in $\Omega_f$ and the rigid-body velocity in $\Omega_s$ as
\begin{equation}
\label{eq:velocity_acceptance}
\widetilde{\bm v}_f^{N+1} =
\begin{cases}
\widetilde{\bm v}_f^{**(k)}, & \bm x\in\Omega_f,\\[1mm]
\bm v_s^{N+1(k)}, & \bm x\in\Omega_s .
\end{cases}
\end{equation}
Equation~\eqref{eq:velocity_acceptance} imposes the rigid-body constraint in the accepted velocity field. Unlike the feedback correction returned between iterations (Section~\ref{subsec:corrector2}), this is a final velocity replacement; its single-pass interpretation is discussed in Section~\ref{subsec:pms_relation}. For a converged step, the fictitious-fluid and rigid-body velocities agree to the prescribed residual tolerance. A nonzero mismatch may introduce divergence errors through SPH operators crossing the interface, which vanish in the zero-residual limit as $\widetilde{\bm v}_f^{**(k)}\rightarrow\widetilde{\bm v}_f^{N+1}$.

\paragraph{Update of particle configuration}

The particle configuration is kept fixed throughout the coupling iterations. The extended-fluid variables and rigid-body velocities are recomputed during the iteration, while particle positions and rigid-body orientation are not advanced. After completion of the coupling procedure, the particle configuration is advanced once to the next physical time level.

The center-of-mass position is updated as
\begin{equation}
\label{eq:rigid_position_update}
\bm X_s^{N+1} = \bm X_s^N + \Delta t\,\bm V_s^{N+1}.
\end{equation}
Let $\bm q_s$ denote the unit quaternion representing the rigid-body orientation. The orientation is updated from the accepted angular velocity using quaternion-based rotational integration~\cite{Asai-Li_ISPH_DEM_ETI_2021}, and the corresponding rotation matrix is denoted by $\mathbf A^{N+1}$. The positions of the constituent rigid-body particles are reconstructed as
\begin{equation}
\label{eq:rigid_particle_position_update}
\bm x_i^{N+1} = \bm X_s^{N+1} + \mathbf A^{N+1} \left( \bm x_i^0-\bm X_s^0 \right), \qquad i\in\mathcal P_s .
\end{equation}
The spatial inertia tensor at the new physical time level is obtained as
\begin{equation}
\label{eq:inertia_update}
\mathbf I_s^{N+1} = \mathbf A^{N+1} \mathbf I_s^0 \left( \mathbf A^{N+1} \right)^{\top}.
\end{equation}
The physical-fluid particles are advected using the ALE transport velocity,
\begin{equation}
\label{eq:fluid_particle_position_update}
\bm x_i^{N+1} = \bm x_i^N + \Delta t\,\bm w_i^{N+1}, \qquad i\in\mathcal P_f .
\end{equation}


\subsection{Relationship between PMS and FD-SPH from a fractional-step perspective}
\label{subsec:pms_relation}

The present partitioned iterative FD-SPH formulation reduces to single-pass coupling when $K_{\max}=1$. In this limit, the extended-fluid predictor and pressure projection are followed by hydrodynamic-load evaluation and the Newton--Euler rigid-body update. The resulting rigid-body velocity is then imposed in the solid domain according to Eq.~\eqref{eq:velocity_acceptance}. This final velocity replacement can be interpreted as a direct-forcing-type correction within a fractional-step framework~\cite{Fadlun_DirectForcing_IB_2000,Uhlmann_DirectForcing_IB_2005}, with the equivalent force density
\begin{equation}
\label{eq:fd_direct_forcing_density}
\bm\lambda_i^{\mathrm{DF}}
=
\frac{\rho_f}{\Delta t}
\left(
\widetilde{\bm v}_{f,i}^{**(1)}
-
\bm v_{s,i}^{N+1(1)}
\right)
=
\frac{\rho_f}{\Delta t}\bm R_i^{(1)},
\qquad i\in\mathcal P_s .
\end{equation}
This expression represents the final velocity correction, rather than an additional force applied during the pressure projection or the hydrodynamic load used in the Newton--Euler equations.

A similar fractional-step structure is employed in the conventional Passively Moving Solid (PMS) method~\cite{Koshizuka_MPS_PMS_origin_1998}. In PMS, solid particles first undergo a fluid-like velocity update, after which their velocities are projected onto rigid-body motion while preserving their total linear and angular momenta. Using the same solid-particle mass and inertia representations as FD-SPH, the PMS projection and its simplified form for equal particle masses, $m_{s,i}=m_s$, are given by
\begin{equation}
\label{eq:pms_mass_weighted_translation}
\bm V_s^{\mathrm{PMS}}
=
\frac{1}{M_s}
\sum_{i\in\mathcal P_s}
m_{s,i}\bm v_{f,i}^{\mathrm{PMS}}
=
\frac{1}{N_s}
\sum_{i\in\mathcal P_s}
\bm v_{f,i}^{\mathrm{PMS}} ,
\end{equation}
\begin{equation}
\label{eq:pms_mass_weighted_rotation}
\bm\omega_s^{\mathrm{PMS}}
=
(\mathbf I_s^N)^{-1}
\sum_{i\in\mathcal P_s}
m_{s,i}\bm r_i^N\times\bm v_{f,i}^{\mathrm{PMS}}
=
\left[
\sum_{i\in\mathcal P_s}
\left\{
(\bm r_i^N\cdot\bm r_i^N)\mathbf 1
-\bm r_i^N\otimes\bm r_i^N
\right\}
\right]^{-1}
\sum_{i\in\mathcal P_s}
\bm r_i^N\times\bm v_{f,i}^{\mathrm{PMS}} ,
\end{equation}
where $\bm v_{f,i}^{\mathrm{PMS}}$ denotes the particle velocity before rigid-body projection, $N_s$ is the number of solid particles, $\sum_{i\in\mathcal P_s}m_{s,i}\bm r_i^N=\bm0$, and $\mathbf I_s^N$ is assumed to be nonsingular. The resulting rigid-body velocity is given by $\bm v_{s,i}^{\mathrm{PMS}}=\bm V_s^{\mathrm{PMS}}+\bm\omega_s^{\mathrm{PMS}}\times\bm r_i^N$.

Thus, PMS and single-pass FD-SPH share a common fractional-step structure and a physical foundation in linear and angular momentum balance. PMS obtains rigid-body motion by projecting fluid-like particle velocities while preserving their momenta, whereas FD-SPH updates rigid-body motion through the Newton--Euler equations using hydrodynamic forces and torques. For the same initial rigid-body state, mass, and inertia tensor, the corresponding discrete momentum conditions can be written as
\begin{equation}
\label{eq:pms_fd_linear_momentum_condition}
\sum_{i\in\mathcal P_s}m_{s,i}\bm v_{f,i}^{\mathrm{PMS}}
=
M_s\bm V_s^N
+\Delta t\left(M_s\bm g+\bm F_f^{N+1(1)}\right),
\end{equation}
\begin{equation}
\label{eq:pms_fd_angular_momentum_condition}
\sum_{i\in\mathcal P_s}m_{s,i}\bm r_i^N\times\bm v_{f,i}^{\mathrm{PMS}}
=
\mathbf I_s^N\bm\omega_s^N
+\Delta t\left[\bm T_f^{N+1(1)}-\bm\omega_s^N\times\left(\mathbf I_s^N\bm\omega_s^N\right)\right].
\end{equation}
These conditions correspond to the translational and rotational updates adopted in Eqs.~\eqref{eq:rigid_translational_update} and~\eqref{eq:rigid_rotational_update}, including the gyroscopic term. Substituting them into the PMS projection in Eqs.~\eqref{eq:pms_mass_weighted_translation} and~\eqref{eq:pms_mass_weighted_rotation} gives the same translational and angular velocities as the Newton--Euler update, before prescribed motion constraints are applied. This establishes their mathematical correspondence under consistent discrete momentum transfer, without requiring the complete coupling algorithms to be identical.

For $K_{\max}>1$, the present FD-SPH formulation extends the single-pass procedure by feeding the kinematic residual back to the preceding fluid velocity iterate through Eq.~\eqref{eq:coupling_velocity_update}. The fluid pressure, hydrodynamic loads, and rigid-body motion are then iteratively updated within the same physical time step, following the strongly coupled FD-SPH framework of Omura et al.~\cite{Omura_Tsuji_FD-SPH_2026}.


\subsection{Overall partitioned iterative algorithm}
\label{subsec:overall_algorithm}

The complete numerical procedure is summarized in Algorithm~\ref{alg:FDSPH}. At the beginning of each physical time step, the velocity iterate is initialized by $\widetilde{\bm v}_f^{N+1(0)}=\widetilde{\bm v}_f^N$. At each coupling iteration, the predictor and incompressibility projection are evaluated, followed by the hydrodynamic-force evaluation and rigid-body update. The resulting kinematic residual is then used either to terminate the coupling procedure or, when another iteration is required, to update the velocity supplied to the subsequent fluid calculation. For multiple rigid bodies, the hydrodynamic loads and Newton--Euler updates are evaluated separately for each body, while the kinematic residuals are assembled into a single global vector for convergence assessment and acceleration.

\begin{algorithm}[tb]
\footnotesize
\caption{Partitioned iterative FD-SPH algorithm}
\label{alg:FDSPH}
\begin{algorithmic}[1]

\Require $\widetilde{\bm v}_f^N$,
$(\bm X_s^N,\bm V_s^N,\bm\omega_s^N,\bm q_s^N)$,
$\Delta t$, $K_{\max}$, $\varepsilon_{\rm rel}$,
$\mathcal{R}_{\mathrm{floor}}$, acceleration strategy

\Ensure $\widetilde{\bm v}_f^{N+1}$,
$\widetilde p^{N+1}$,
$(\bm X_s^{N+1},\bm V_s^{N+1},
\bm\omega_s^{N+1},\bm q_s^{N+1})$

\State Initialize
$\widetilde{\bm v}_f^{N+1(0)}
\leftarrow\widetilde{\bm v}_f^N$
\State Clear the acceleration history for the new physical time step

\For{$k=1,\ldots,K_{\max}$}

    \State \textbf{Predictor:}
    compute $\widetilde{\bm v}_f^{*(k)}$
    using Eq.~\eqref{eq:predictor_step}

    \State \textbf{Corrector-1:}
    solve Eq.~\eqref{eq:pressure_poisson}
    for $\widetilde p^{N+1(k)}$

    \State Compute $\widetilde{\bm v}_f^{**(k)}$
    using Eq.~\eqref{eq:pressure_projection}

    \State \textbf{Hydrodynamic loads:}
    compute $\bm F_f^{N+1(k)}$
    and $\bm T_f^{N+1(k)}$

    \State \textbf{Rigid-body update:}
    compute $\bm V_s^{N+1(k)}$
    and $\bm\omega_s^{N+1(k)}$
    using Eqs.~\eqref{eq:rigid_translational_update}
    and~\eqref{eq:rigid_rotational_update}

    \State Apply prescribed motion constraints, if any,
    and compute $\bm v_s^{N+1(k)}$
    using Eq.~\eqref{eq:rigid_velocity_field_disc}

    \State Compute
    $\bm R_i^{(k)}
    =\widetilde{\bm v}_{f,i}^{**(k)}
    -\bm v_{s,i}^{N+1(k)}$
    for $i\in\mathcal P_s$

    \State Compute $\mathcal{R}^{(k)}$
    using Eq.~\eqref{eq:residual_norm}

    \If{$K_{\max}=1$}
        \State Terminate after the prescribed single pass
        \State \textbf{break}
    \EndIf

    \If{$k=1$}
        \If{$\mathcal{R}^{(1)}
        \le\mathcal{R}_{\mathrm{floor}}$}
            \State Accept the current coupled state
            as numerically converged
            \State \textbf{break}
        \EndIf
        \State Set
        $\widehat{\mathcal{R}}^{(1)}\leftarrow 1$
    \Else
        \State Compute
        $\widehat{\mathcal{R}}^{(k)}
        =\mathcal{R}^{(k)}/\mathcal{R}^{(1)}$
        using Eq.~\eqref{eq:relative_residual}

        \If{$\widehat{\mathcal{R}}^{(k)}
        \le\varepsilon_{\rm rel}$}
            \State Accept the current coupled state as converged
            \State \textbf{break}
        \EndIf
    \EndIf

    \If{$k=K_{\max}$}
        \State Retain the current state
        as a nonconverged approximation
        \State \textbf{break}
    \EndIf

    \State Assemble $\bm u^{(k)}$
    and $\bm r_{\mathrm{cpl}}^{(k)}$
    and update the acceleration history

    \State Construct $\bm C^{(k)}$
    using the selected acceleration strategy

    \Statex \hspace{\algorithmicindent}
    \emph{For Anderson, use $\omega_0=0.2$
    initially and as a fallback.}

    \State Update
    $\widetilde{\bm v}_f^{N+1(k)}$
    using Eq.~\eqref{eq:coupling_velocity_update}

    \State Restore the rigid-body working variables
    to their accepted values at $t^N$

\EndFor

\State Set
$\widetilde p^{N+1}\leftarrow\widetilde p^{N+1(k)}$,
$\bm V_s^{N+1}\leftarrow\bm V_s^{N+1(k)}$,
and $\bm\omega_s^{N+1}\leftarrow\bm\omega_s^{N+1(k)}$

\State Construct the accepted velocity
$\widetilde{\bm v}_f^{N+1}$
using Eq.~\eqref{eq:velocity_acceptance}

\State Update the rigid-body and physical-fluid
particle configurations

\end{algorithmic}
\end{algorithm}
\FloatBarrier


\section{Numerical assessment and experimental validation}
\label{sec:numerical_examples}

Four problems are considered to assess complementary aspects of the present rigid-body-specific FD-SPH formulation. The two-dimensional settling-disk benchmark first examines agreement with established numerical results in a case where weak and strong coupling produce similar responses. The buoyancy-driven motion of a lightweight cubic body then examines coupling stability, equilibrium configurations, and convergence of the partitioned iteration. Wet-bed dam-break experiments involving a floating rectangular body provide experimental validation of the predicted free-surface and rigid-body motions in two and three dimensions. Finally, a two-dimensional wet-bed dam-break problem with two lightweight star-shaped rigid bodies provides a numerical demonstration of the applicability of the same formulation to multiple rigid bodies with complex geometries featuring sharp convex and re-entrant corners, together with a dynamically varying inter-body flow.

Throughout this section and in the figure labels, weak coupling denotes the single-pass FD-SPH configuration with $K_{\max}=1$, whereas strong coupling denotes the within-step partitioned iterative configuration when the relative-residual criterion is satisfied. Unless otherwise stated, the strong-coupling calculations use $K_{\max}=2000$ together with $\varepsilon_{\rm rel}=10^{-3}$ in Eq.~\eqref{eq:convergence}. The value of $K_{\max}$ serves as a safeguard against an indefinitely continuing coupling iteration and is not used to define convergence. For $K_{\max}>1$, a physical time step is classified as a converged strong-coupling step only when the relative-residual criterion is satisfied; a step reaching $K_{\max}$ without satisfying this criterion is retained as a nonconverged maximum-iteration approximation. Numerical stability is therefore assessed separately from coupling convergence. For all production strong-coupling simulations reported below, the prescribed relative-residual criterion was satisfied at every physical time step before reaching $K_{\max}$. This statement concerns the production simulations and excludes the deliberately nonconvergent unrelaxed-iteration comparison used to assess acceleration strategies.


\subsection{Numerical benchmark assessment: two-dimensional disk sedimentation}
\label{subsec:fundamental_assessment}

The settling of a circular disk in a viscous fluid is considered using the benchmark configuration reported by Glowinski et al.~\cite{Glowinski_DFFD_2001} and Wan and Turek~\cite{Wan_Turek_FBM_2006}. The two-dimensional channel has a width of $2~\mathrm{cm}$ and a height of $6~\mathrm{cm}$. A disk of diameter $0.25~\mathrm{cm}$ is initially centered at $(1.0,4.0)~\mathrm{cm}$. The fluid and disk densities are $\rho_f=1.0~\mathrm{g/cm^3}$ and $\rho_s=1.25~\mathrm{g/cm^3}$, respectively, and the fluid kinematic viscosity is $\nu_f=0.1~\mathrm{cm^2/s}$. The disk and fluid are initially at rest.

\begin{figure}[tbp]
    \centering
    \includegraphics[width=\linewidth]{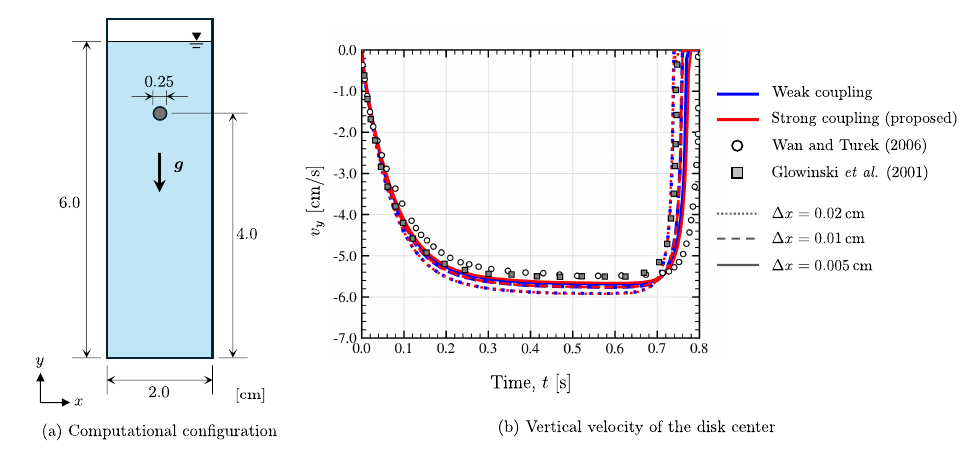}
    \caption{
    Two-dimensional settling-disk benchmark: (a) computational configuration and (b) vertical velocity under weak and strong coupling at three particle spacings. Reference data were digitized from Glowinski et al.~\cite{Glowinski_DFFD_2001} and Wan and Turek~\cite{Wan_Turek_FBM_2006}.
    }
    \label{fig:settling_disk}
\end{figure}

The present SPH calculations use three particle spacings, $\Delta x=0.02$, $0.01$, and $0.005~\mathrm{cm}$, with a fixed time increment of $\Delta t=5.0\times10^{-5}~\mathrm{s}$. Figure~\ref{fig:settling_disk}(a) shows the computational configuration, and Fig.~\ref{fig:settling_disk}(b) compares the predicted vertical disk velocity with reference histories digitized from the published figures of Glowinski et al.~\cite{Glowinski_DFFD_2001} and Wan and Turek~\cite{Wan_Turek_FBM_2006}.

The present calculations reproduce the overall settling behavior and characteristic velocity history reported in the reference studies. As the particle spacing is reduced, the predicted velocity histories show progressively closer agreement with the published numerical results. Some difference from the reference calculations remains even at the finest resolution. At each particle spacing, however, both the weak- and strong-coupling calculations remain stable and produce nearly identical velocity histories, indicating that within-step iteration has little influence on the solution for this benchmark under the tested conditions.

This benchmark provides a baseline assessment of the present implementation against established numerical results. The following lightweight-body tests examine conditions in which weak coupling becomes unstable and demonstrate the benefit of strong coupling.

\subsection{Buoyancy-driven motion of a lightweight cubic body}
\label{subsec:floating_cube}

A three-dimensional cubic body with side length $L=50~\mathrm{cm}$ is initially at rest in an incompressible Newtonian liquid, with its upper face coincident with the initial free surface. The computational tank occupies $(0,100)\times(0,150)\times(0,100)~\mathrm{cm}$ in the $(x,y,z)$ directions. The initial water depth is $75~\mathrm{cm}$, so the cube is fully submerged with its bottom face at $y=25~\mathrm{cm}$. Gravity acts in the negative $y$ direction with magnitude $g=981~\mathrm{cm/s^2}$. The fluid density and kinematic viscosity are $\rho_f=1.0~\mathrm{g/cm^3}$ and $\nu_f=1.0~\mathrm{cm^2/s}$, respectively. The specific gravity is defined as $G_s=\rho_s/\rho_f$.

The heave-only calculations cover $G_s=0.1$--$1.0$ in increments of $0.1$, while the heave-and-rotation configurations presented below use $G_s=0.1$, $0.2$, $0.25$, $0.5$, $0.75$, $0.8$, and $0.9$. Unless otherwise stated, the particle spacing and time increment are $\Delta x=1.25~\mathrm{cm}$ and $\Delta t=5.0\times10^{-4}~\mathrm{s}$, respectively. Additional calculations with smaller time increments are performed only to examine the time-step sensitivity of the weak-coupling instability.

\begin{figure}[tbp]
    \centering
    \includegraphics[]{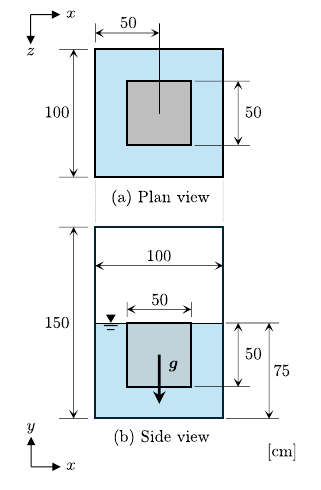}
    \caption{Three-dimensional configuration of the buoyancy-driven cube test: (a) plan view and (b) side view. Dimensions are in centimeters.}
    \label{fig:floating_cube_model}
\end{figure}

Two motion constraints are considered: heave only and heave with three-dimensional rotation. Horizontal translation is constrained in both cases. Figure~\ref{fig:floating_cube_model} shows the initial configuration and computational domain.

\subsubsection{Heave-only response and long-time stability}
\label{subsubsec:cube_heave}

With rotation suppressed, hydrostatic equilibrium requires $\rho_s V_s=\rho_f V_{\rm sub}$, giving $V_{\rm sub}/V_s=G_s$. For an upright cube with $0<G_s<1$, the theoretical equilibrium freeboard, defined as the elevation of the upper face relative to the equilibrium free surface, is $h_{\rm top,eq}=(1-G_s)L$.

The plotted height is measured from the tank bottom. Accounting for the change in free-surface elevation due to water-volume conservation in the finite tank, the equilibrium elevation of the cube's upper face is
\begin{equation}
    y_{\rm top,eq}
    =
    H_{\rm w,0}
    -\frac{(1-G_s)V_s}{A_T}
    +(1-G_s)L
    =
    (112.5-37.5G_s)~\mathrm{cm},
    \label{eq:cube_equilibrium_height}
\end{equation}
where $H_{\rm w,0}=75~\mathrm{cm}$ is the initial water depth, $A_T=100\times100~\mathrm{cm^2}$ is the tank cross-sectional area, and $V_s=L^3$ with $L=50~\mathrm{cm}$.

\begin{figure}[tbp]
    \centering
    \includegraphics[width=0.95\linewidth]{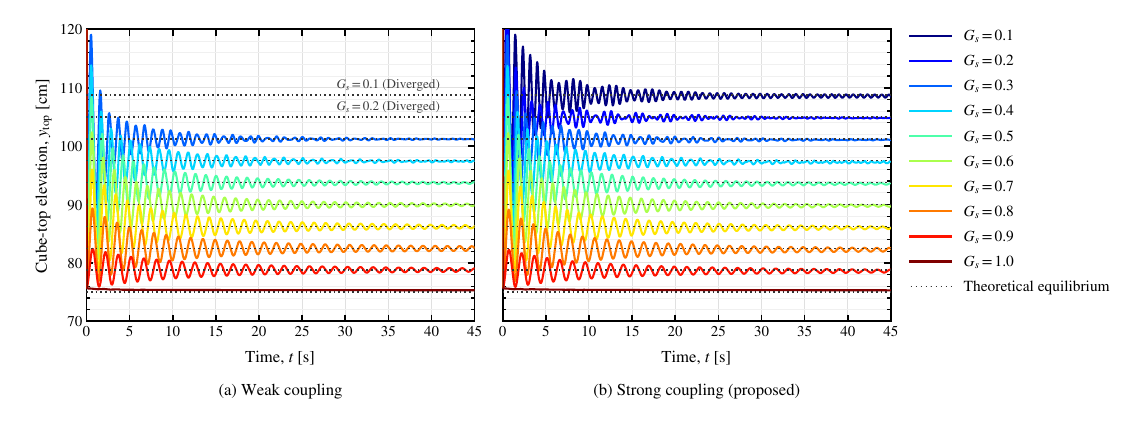}
    \caption{Heave-only responses for representative specific gravities under (a) weak and (b) strong coupling. Dotted lines indicate the theoretical equilibrium heights from Eq.~\eqref{eq:cube_equilibrium_height}.}
    \label{fig:cube_heave_history}
\end{figure}

Figure~\ref{fig:cube_heave_history} compares the heave responses with the theoretical equilibrium heights. The weak-coupling solutions remain stable for $G_s=0.3$--$1.0$, and the buoyant cases in this range approach the values given by Eq.~\eqref{eq:cube_equilibrium_height}. For $G_s=0.1$ and $0.2$, however, the weak-coupling solutions exhibit unphysical upward acceleration immediately after release. In contrast, the strong-coupling solutions remain stable for all tested density ratios and approach the corresponding hydrostatic equilibrium states.

\begin{figure}[tbp]
    \centering
    \includegraphics[width=0.95\linewidth]{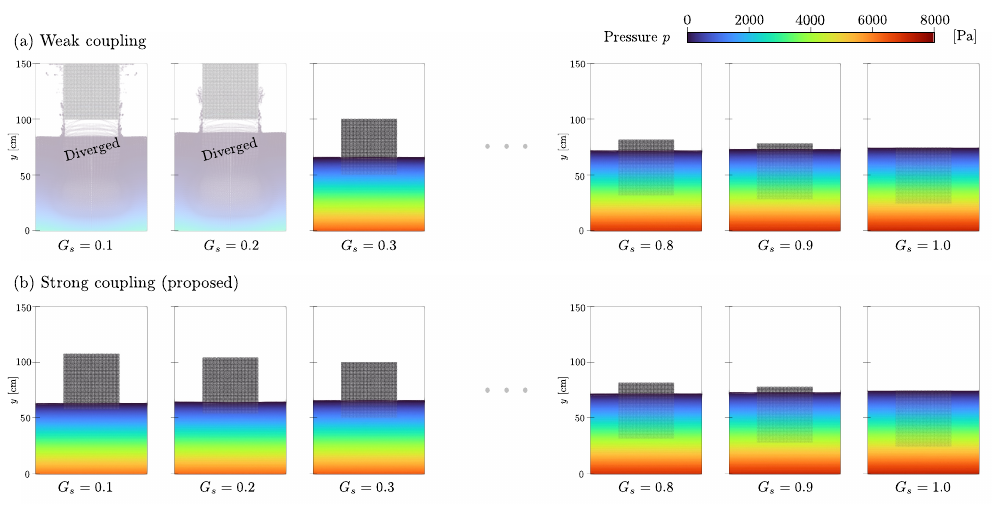}
    \caption{Pressure distributions and rigid-body positions under (a) weak and (b) strong coupling. Stable cases are shown at $t=45~\mathrm{s}$, whereas the unstable weak-coupling cases with $G_s=0.1$ and $0.2$ are shown at $t=0.1~\mathrm{s}$.}
    \label{fig:cube_heave_pressure}
\end{figure}

Figure~\ref{fig:cube_heave_pressure} shows the pressure distributions and rigid-body positions for the stable cases at $t=45~\mathrm{s}$, together with representative snapshots of the divergent weak-coupling cases. The weak-coupling cases with $G_s=0.1$ and $0.2$ exhibit excessive upward displacement, whereas the strong-coupling cases maintain stable configurations. The neutrally buoyant body with $G_s=1.0$ remains close to its initial position.

\subsubsection{Weak-coupling instability}
\label{subsubsec:cube_coupling_behavior}

To examine whether the weak-coupling failure is primarily a consequence of the selected time increment, weak-coupling calculations for $G_s=0.1$ were performed with $\Delta t=5.0\times10^{-4}$, $1.0\times10^{-4}$, $1.0\times10^{-5}$, and $1.0\times10^{-6}~\mathrm{s}$, while the spatial discretization and other numerical settings were kept unchanged. In all four cases, numerical instability developed within the first ten physical time steps and the body exhibited unphysical upward acceleration. Thus, reducing the time increment by a factor of up to 500 did not suppress the instability, indicating that the observed failure is not solely a large-time-step artifact.

\begin{figure*}[tbp]
    \centering
    \includegraphics[width=0.95\linewidth]{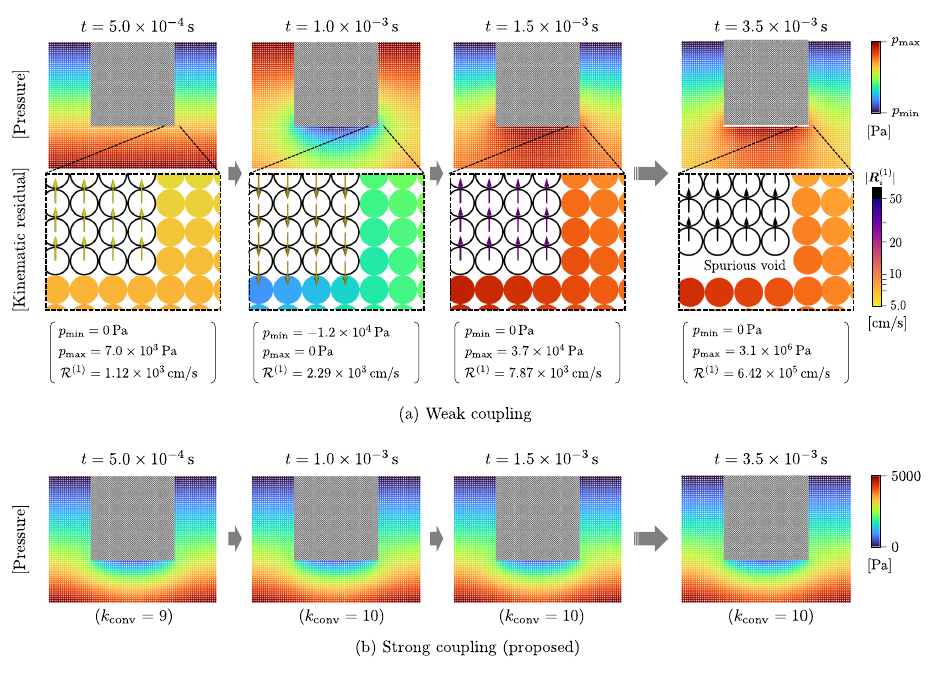}
    \caption{Early development of the weak-coupling instability for the heave-only cube with $G_s=0.1$. The upper and lower rows in (a) show pressure and kinematic-residual vectors, respectively. Vector lengths are normalized for direction only, whereas colors indicate $\lvert\bm R_i^{(1)}\rvert$ on a logarithmic scale. Pressure color scales are set separately for each snapshot; $p_{\min}$, $p_{\max}$, and $\mathcal{R}^{(1)}$ are indicated.}
    \label{fig:cube_weak_instability}
\end{figure*}

Figure~\ref{fig:cube_weak_instability} examines the early evolution of a representative weak-coupling calculation. The pressure anomaly beneath the body alternates in sign between successive physical time steps. At the same time, the particle-wise kinematic-residual vectors within the rigid-body domain reverse direction while their magnitude increases. By the seventh physical time step, a spurious particle-free gap also develops beneath the body. Because the single-pass formulation advances the physical solution without reducing this fluid--rigid-body mismatch, the unresolved mismatch is carried into the subsequent physical step and undergoes an oscillatory step-to-step amplification.

Together with the persistence of the instability over the four tested time increments, these observations are consistent with an oscillatory partitioned-coupling instability rather than a failure attributable solely to an excessively large time increment. The present results do not constitute an eigenvalue-based stability analysis, but they directly show that the unstable response is accompanied by alternating pressure states and a growing kinematic mismatch. In the following subsection, the corresponding strong-coupling behavior is assessed quantitatively through within-step residual convergence and full-run convergence statistics.

\subsubsection{Convergence acceleration and full-run behavior}
\label{subsubsec:cube_acceleration}

The convergence behavior of the within-step partitioned iteration is first examined at the initial physical time step of the heave-only $G_s=0.1$ case. Static relaxation with $\omega=1.0$, $0.5$, and $0.2$, Aitken dynamic relaxation, and Anderson acceleration with $m=3$ are compared using the same initial physical state, spatial discretization, and time increment. A common relative-residual tolerance of $10^{-3}$ is prescribed.

\begin{figure}[tbp]
    \centering
    \includegraphics[]{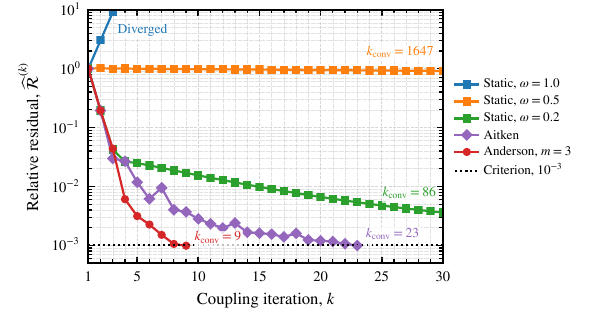}
    \caption{Relative-residual convergence at the initial physical time step for $G_s=0.1$, comparing static relaxation, Aitken relaxation, and Anderson acceleration. Only the first 30 coupling iterations are displayed. The dotted line indicates the tolerance $10^{-3}$, and $k_{\mathrm{conv}}$ denotes the total number of iterations required for convergence.}
    \label{fig:cube_acceleration}
\end{figure}

Figure~\ref{fig:cube_acceleration} shows the relative residual $\widehat{\mathcal{R}}^{(k)}$ against the coupling iteration. Static relaxation with $\omega=1.0$ diverges, whereas $\omega=0.5$ and $0.2$ reach the tolerance in 1647 and 86 iterations, respectively, demonstrating the sensitivity of convergence to the relaxation factor. Aitken relaxation converges in 23 iterations, with some nonmonotonic residual variation, while Anderson acceleration reaches the same tolerance in nine iterations, the fewest among the tested strategies. Anderson acceleration with $m=3$ is therefore adopted for the subsequent strong-coupling calculations unless otherwise stated.

The preceding comparison concerns only the initial physical time step. As a detailed quantitative assessment of the full-run convergence behavior, the complete heave-only simulations with $G_s=0.1$--$1.0$ over $0\le t\le45~\mathrm{s}$ were analyzed. All time steps satisfied the prescribed relative-residual criterion before reaching $K_{\max}=2000$.

\begin{figure*}[tbp]
    \centering
    \includegraphics[width=0.95\linewidth]{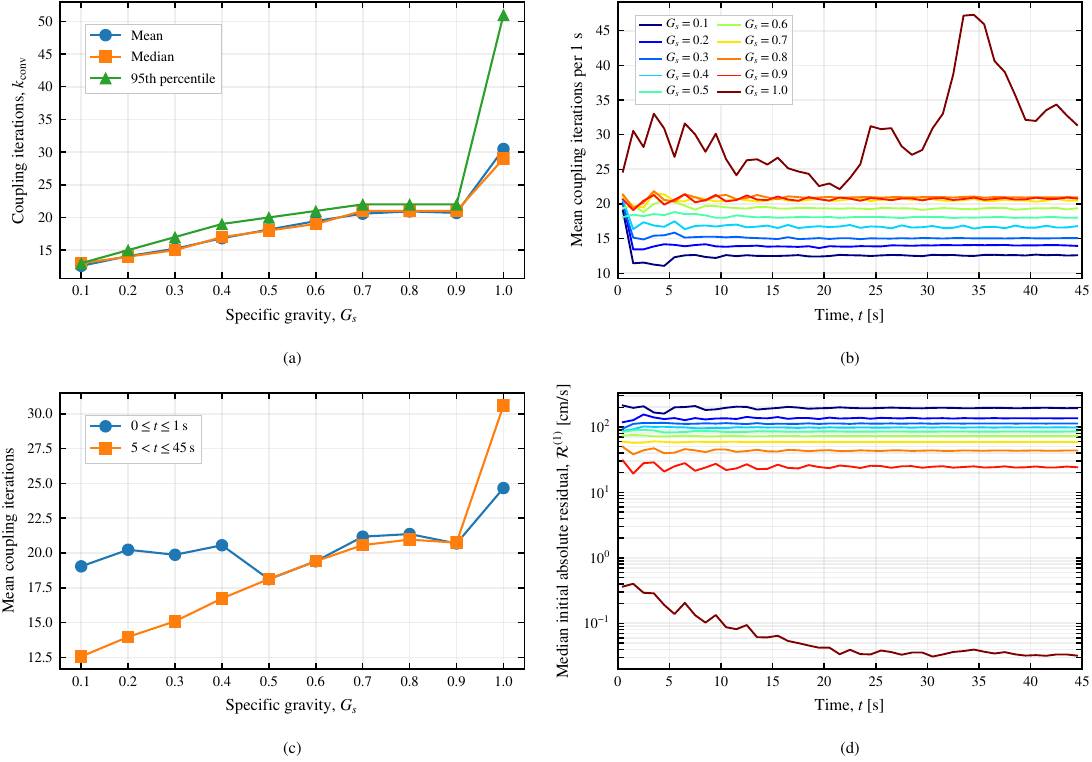}
    \caption{Full-run convergence statistics of Anderson-accelerated strong coupling for $G_s=0.1$--$1.0$: (a) iteration-count statistics, (b) temporal mean iteration count, (c) comparison between initial and later intervals, and (d) median initial absolute residual $\mathcal{R}^{(1)}$.}
    \label{fig:cube_convergence_statistics}
\end{figure*}

Figure~\ref{fig:cube_convergence_statistics}(a) summarizes the mean, median, and 95th-percentile iteration counts. The mean ranges from approximately 12.6 to 30.5, while the 95th percentile remains below 51 for all cases. Figure~\ref{fig:cube_convergence_statistics}(b) shows the corresponding temporal variation using $1~\mathrm{s}$ intervals. For the lightweight cases, the larger iteration counts are concentrated mainly in the initial transient and decrease as the body approaches its floating state. This trend is also evident in Fig.~\ref{fig:cube_convergence_statistics}(c), which compares the mean iteration counts during $0\le t\le1~\mathrm{s}$ and $5<t\le45~\mathrm{s}$.

The $G_s=1.0$ case shows a different trend. As shown in Fig.~\ref{fig:cube_convergence_statistics}(d), its initial absolute residual $\mathcal{R}^{(1)}$ is several orders of magnitude smaller than those of the buoyant cases, consistent with the nearly stationary state of the neutrally buoyant body. Because convergence is defined by $\widehat{\mathcal{R}}^{(k)}\le10^{-3}$, the smaller initial residual requires a correspondingly smaller absolute residual at convergence, accounting, at least in part, for the comparatively larger iteration counts in this case.

Overall, the largest observed iteration count is 127, well below $K_{\max}=2000$, and all analyzed physical time steps satisfy the prescribed convergence criterion. These results demonstrate that convergence is maintained throughout the complete 45~s heave-only simulations, rather than only at the representative initial time step.
Although iterative coupling requires additional fluid solves within each physical time step, it enables stable simulations of lightweight rigid bodies under conditions where single-pass coupling fails, even with substantially reduced time increments. Anderson acceleration further reduces the number of iterations required to achieve convergence, making the additional computational effort more manageable.

\subsubsection{Heave and rotation: density-dependent orientation}
\label{subsubsec:cube_rotation}

\begin{figure}[tbp]
    \centering
    \includegraphics[width=0.95\linewidth]{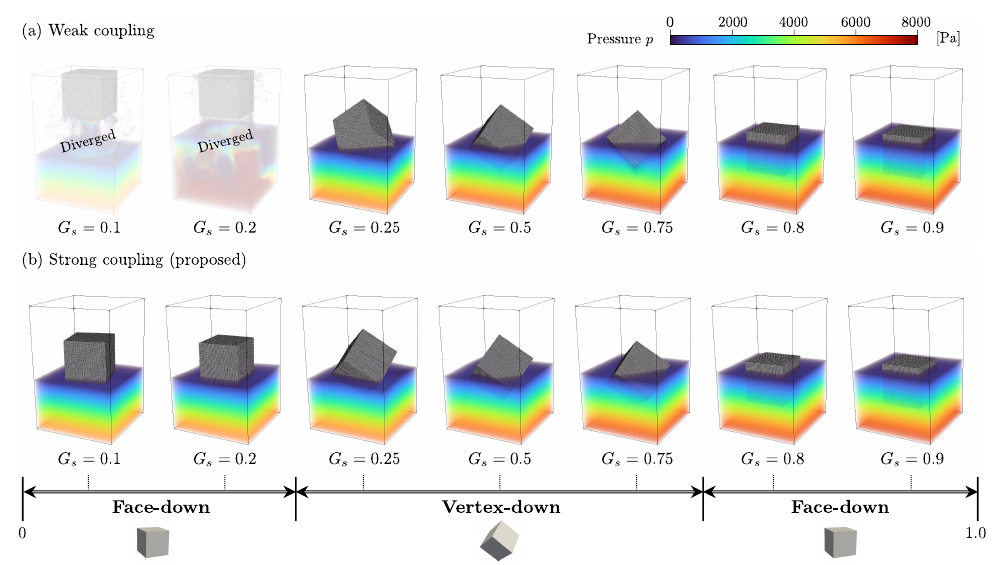}
    \caption{Pressure distributions and cube orientations under (a) weak and (b) strong coupling. Stable cases are shown at $t=30~\mathrm{s}$, whereas the unstable weak-coupling cases with $G_s=0.1$ and $0.2$ are shown at $t=0.1~\mathrm{s}$. The lower schematic summarizes observed orientations at the sampled specific gravities; the divisions do not represent theoretical stability boundaries.}
    \label{fig:cube_orientation}
\end{figure}

The rotational degrees of freedom are released while horizontal translation remains constrained. The calculations start from the same upright and stationary configuration as the heave-only cases, without any imposed initial angular perturbation or angular velocity. Figure~\ref{fig:cube_orientation} shows the pressure distributions and body orientations at $t=30~\mathrm{s}$. Under weak coupling, the cases with $G_s=0.1$ and $0.2$ exhibit unphysical upward motion. Under strong coupling, stable solutions are obtained for all tested specific gravities, with face-down orientations at $G_s=0.1$, $0.2$, $0.8$, and $0.9$ and vertex-down orientations at $G_s=0.25$, $0.5$, and $0.75$.

Gilbert's hydrostatic analysis predicts stable face-down orientations at low and high specific gravities and stable vertex-down orientations over an intermediate range. These stability ranges overlap near $G_s=0.2$ and $0.8$, so the observed face-down orientations at those ratios are consistent with the theory, although other stable orientations are also possible~\cite{Gilbert_HowThingsFloat_1991}.

Morikawa and Asai's three-dimensional fluid--structure interaction simulations likewise reported a face-down orientation at $G_s=0.1$, either face-down or vertex-down orientations at $G_s=0.2$ depending on the initial orientation, and a vertex-down orientation at $G_s=0.7$~\cite{Morikawa_SPH-EISPH_FSI_2021}. Their setup differs from the present one, so this comparison concerns the observed orientations rather than the transient motion.

Strong coupling maintains stable three-dimensional heave and rotation across the tested specific gravities, including the lightweight cases in which weak coupling diverges. The orientations observed at $t=30~\mathrm{s}$ are consistent with stable configurations identified in previous studies.

\subsection{Experimental validation: dam-break flow interacting with a floating body}
\label{subsec:dam_float}

\subsubsection{Experimental configuration and numerical cases}
\label{subsubsec:dam_setup}

\begin{figure}[tbp]
    \centering
    \includegraphics[width=0.95\linewidth]{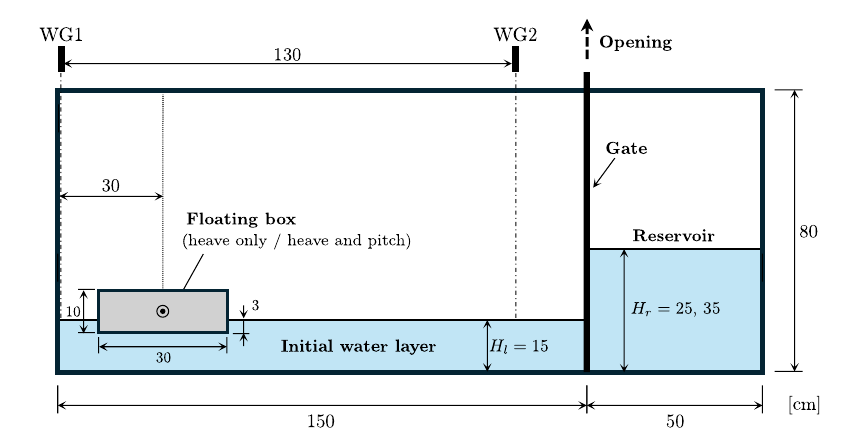}
    \caption{Wet-bed dam-break configuration based on the experiment of Wang and Dong~\cite{Wang-Dong_Exp_dam-floating-box_2022}. Dimensions are in centimeters. Numerical calculations use instantaneous gate release.}
    \label{fig:dam_float_setup}
\end{figure}

The experiments of Wang and Dong~\cite{Wang-Dong_Exp_dam-floating-box_2022} are used to assess the response of a lightweight rectangular floating body to a wet-bed dam-break flow. The experimental tank is $200~\mathrm{cm}$ long, $80~\mathrm{cm}$ high, and $60~\mathrm{cm}$ wide. The gate is located $150~\mathrm{cm}$ from the left wall, leaving a $50~\mathrm{cm}$-long reservoir. WG1 is located at the left wall, and WG2 is located $130~\mathrm{cm}$ from that wall, as shown in Fig.~\ref{fig:dam_float_setup}. The wave elevations at these gauges are denoted by $\eta_1$ and $\eta_2$, respectively.

The experimental floating body measures $30~\mathrm{cm}$ in the streamwise direction, $10~\mathrm{cm}$ in height, and $59.5~\mathrm{cm}$ in the transverse direction. Its experimentally reported mass is $5.35~\mathrm{kg}$, corresponding to a specific gravity of approximately $0.300$; $G_s=0.299$ is prescribed in the present calculations. The initial draft is $3~\mathrm{cm}$, and the initial body center is located at $X_0=30~\mathrm{cm}$ from the left wall. Two motion conditions are considered: heave only, and heave combined with pitch. In the heave-only calculations, rotation is suppressed to approximate the experimentally restrained configuration, in which small pitch motions can occur. Heave is defined as the vertical displacement of the body center of mass from its initial position, with upward displacement taken as positive. Pitch is measured from the initial horizontal orientation, with clockwise rotation defined as positive. The experimental time histories used for comparison were digitized from the published figures of Wang and Dong~\cite{Wang-Dong_Exp_dam-floating-box_2022}.

The initial downstream water depth is $H_l=15~\mathrm{cm}$, and the reservoir depths are $H_r=25$ and $35~\mathrm{cm}$. The fluid density, kinematic viscosity, and gravitational acceleration are $\rho_f=1.0~\mathrm{g/cm^3}$, $\nu_f=0.01~\mathrm{cm^2/s}$, and $g=981~\mathrm{cm/s^2}$, respectively. Two-dimensional calculations are performed at $\Delta x=1.0$ and $0.25~\mathrm{cm}$, whereas the three-dimensional calculations use $\Delta x=1.0~\mathrm{cm}$. In the two-dimensional calculations, the system is interpreted per unit transverse thickness; accordingly, the rigid-body mass and hydrodynamic loads are evaluated per unit transverse width. In the three-dimensional calculations, the transverse body width is set to $58~\mathrm{cm}$ rather than the experimental value of $59.5~\mathrm{cm}$ so that a gap of one particle spacing ($1~\mathrm{cm}$) is retained between the body and each side wall at $\Delta x=1.0~\mathrm{cm}$. The tank width remains $60~\mathrm{cm}$, and the rigid-body mass is determined from the prescribed density and the corresponding numerical body volume. All calculations use $\Delta t=2.0\times10^{-4}~\mathrm{s}$. For both reservoir depths, the two-dimensional calculations consider both weak and strong coupling at each particle spacing and under both motion constraints, whereas the three-dimensional calculations employ strong coupling only.

The physical gate and its lifting mechanism are not modeled; instead, the initially separated water columns are released instantaneously. Wang and Dong assessed their gate-opening times using the sudden-release criterion $t_{\rm open}\sqrt{g/H_r}<\sqrt{2}$ and reported that the selected cases satisfy this criterion~\cite{Lauber_Hager_DamBreak_1998,Wang-Dong_Exp_dam-floating-box_2022}. Instantaneous release is therefore adopted as an approximation of the rapid experimental gate opening.

\subsubsection{Heave-only response}
\label{subsubsec:dam_heave}

\begin{figure}[tbp]
    \centering
    \includegraphics[width=0.9\linewidth]{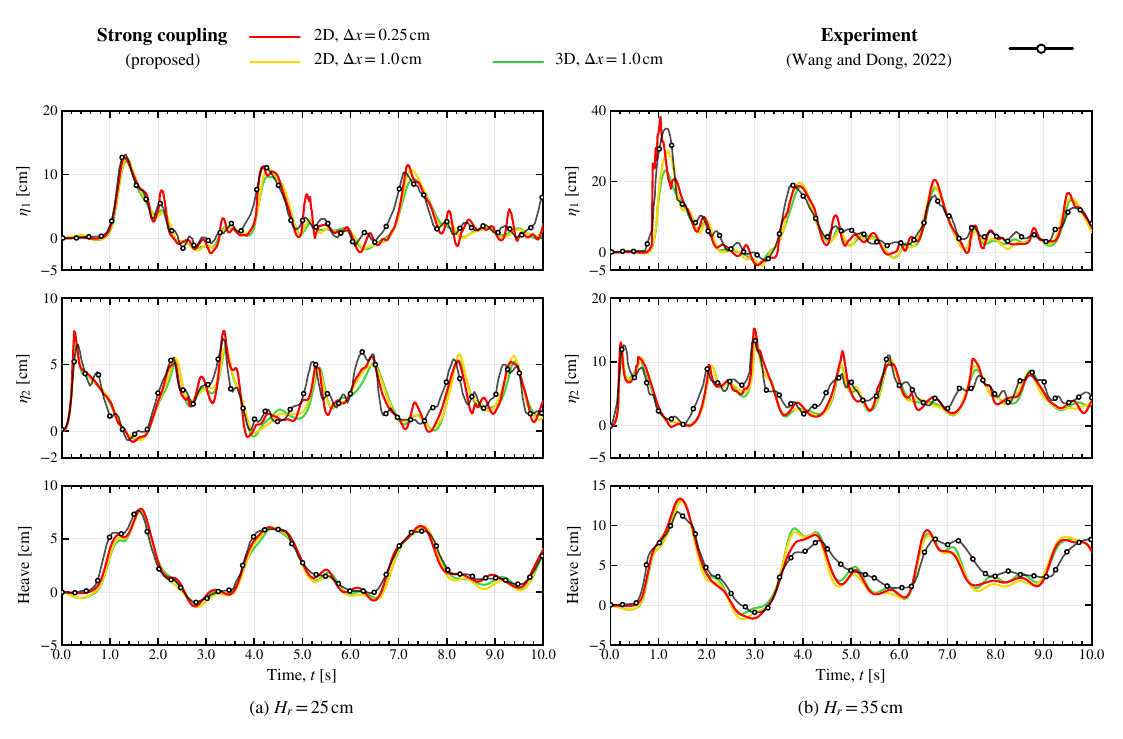}
    \caption{Heave-only responses compared with digitized experimental data from Wang and Dong [35]. (a) $H_r=25$ cm and (b) $H_r=35$ cm. Rows show wave elevations at WG1 and WG2, and body heave. Here, $\eta_1$ and $\eta_2$ denote the wave elevations at WG1 and WG2, respectively.}
    \label{fig:dam_heave_history}
\end{figure}

Figure~\ref{fig:dam_heave_history} compares the strong-coupling predictions of wave elevation at WG1 and WG2 and body heave with the experimental histories of Wang and Dong~\cite{Wang-Dong_Exp_dam-floating-box_2022}. For both reservoir depths, the two-dimensional calculations at $\Delta x=1.0$ and $0.25~\mathrm{cm}$ and the three-dimensional calculation at $\Delta x=1.0~\mathrm{cm}$ reproduce the principal oscillations and their overall timing. Agreement with the experimental measurements is closer for $H_r=25~\mathrm{cm}$; for $H_r=35~\mathrm{cm}$, noticeable differences remain in the first WG1 crest and some subsequent heave extrema. Nevertheless, the main sequence of wave arrival and body oscillation is captured at both reservoir depths.

\begin{figure}[tbp]
    \centering
    \includegraphics[width=0.9\linewidth]{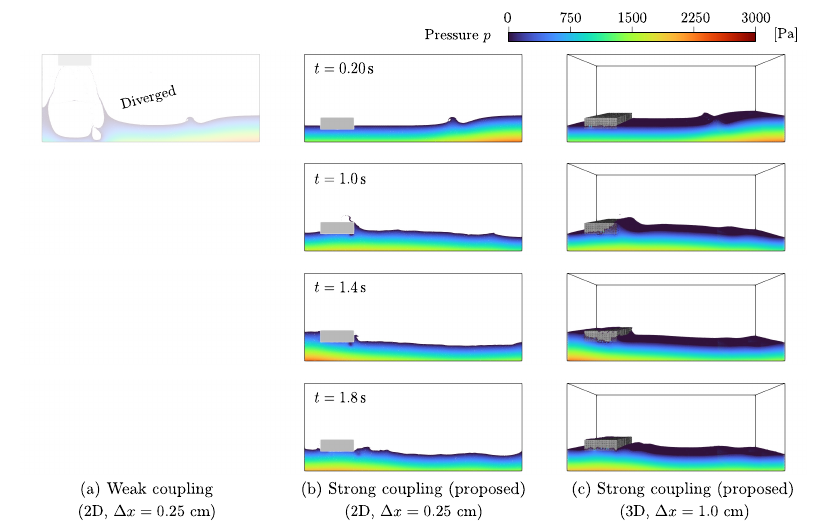}
    \caption{Heave-only snapshots for $H_r=25~\mathrm{cm}$: (a) weak coupling in 2D, (b) strong coupling in 2D, and (c) strong coupling in 3D. Rows show the indicated times; colors denote pressure in Pa. Weak-coupling results are shown only at the first snapshot.}
    \label{fig:dam_heave_Hr25}
\end{figure}

\begin{figure}[tbp]
    \centering
    \includegraphics[width=0.9\linewidth]{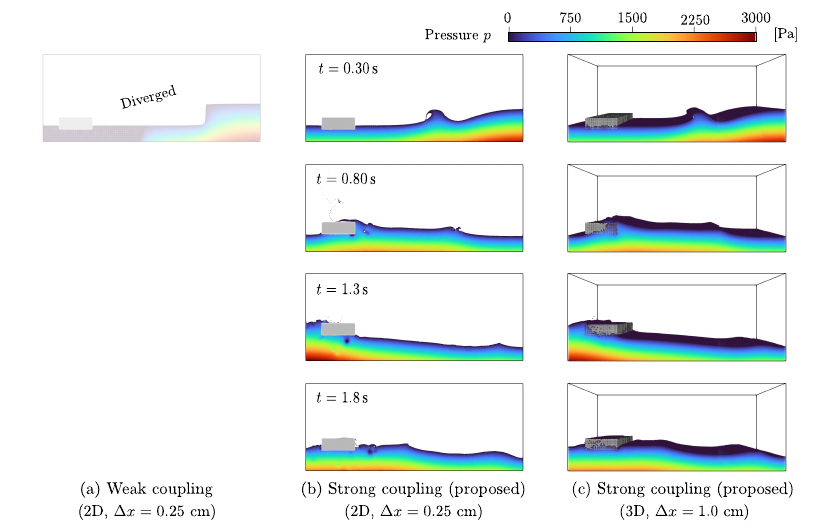}
    \caption{Heave-only snapshots for $H_r=35~\mathrm{cm}$: (a) weak coupling in 2D, (b) strong coupling in 2D, and (c) strong coupling in 3D. Rows show the indicated times; colors denote pressure in Pa. Weak-coupling results are shown only at the first snapshot.}
    \label{fig:dam_heave_Hr35}
\end{figure}

At both two-dimensional particle spacings, weak coupling produces an unphysical body response before bore arrival, whereas strong coupling remains stable throughout the simulated interval. Figures~\ref{fig:dam_heave_Hr25} and \ref{fig:dam_heave_Hr35} illustrate the early weak-coupling instability at $\Delta x=0.25~\mathrm{cm}$ and the corresponding strong-coupling flow and body responses. The two- and three-dimensional strong-coupling results exhibit similar overall flow and body responses. The remaining discrepancies may partly reflect the idealized instantaneous gate release and, in the three-dimensional calculations, the modified body width and limited resolution of the narrow side-wall gaps. These modeling differences should be considered when interpreting the quantitative agreement with the experiments, particularly for local three-dimensional flow effects.

\subsubsection{Coupled heave-and-pitch response}
\label{subsubsec:dam_pitch}

\begin{figure}[tb]
    \centering
    \includegraphics[width=0.9\linewidth]{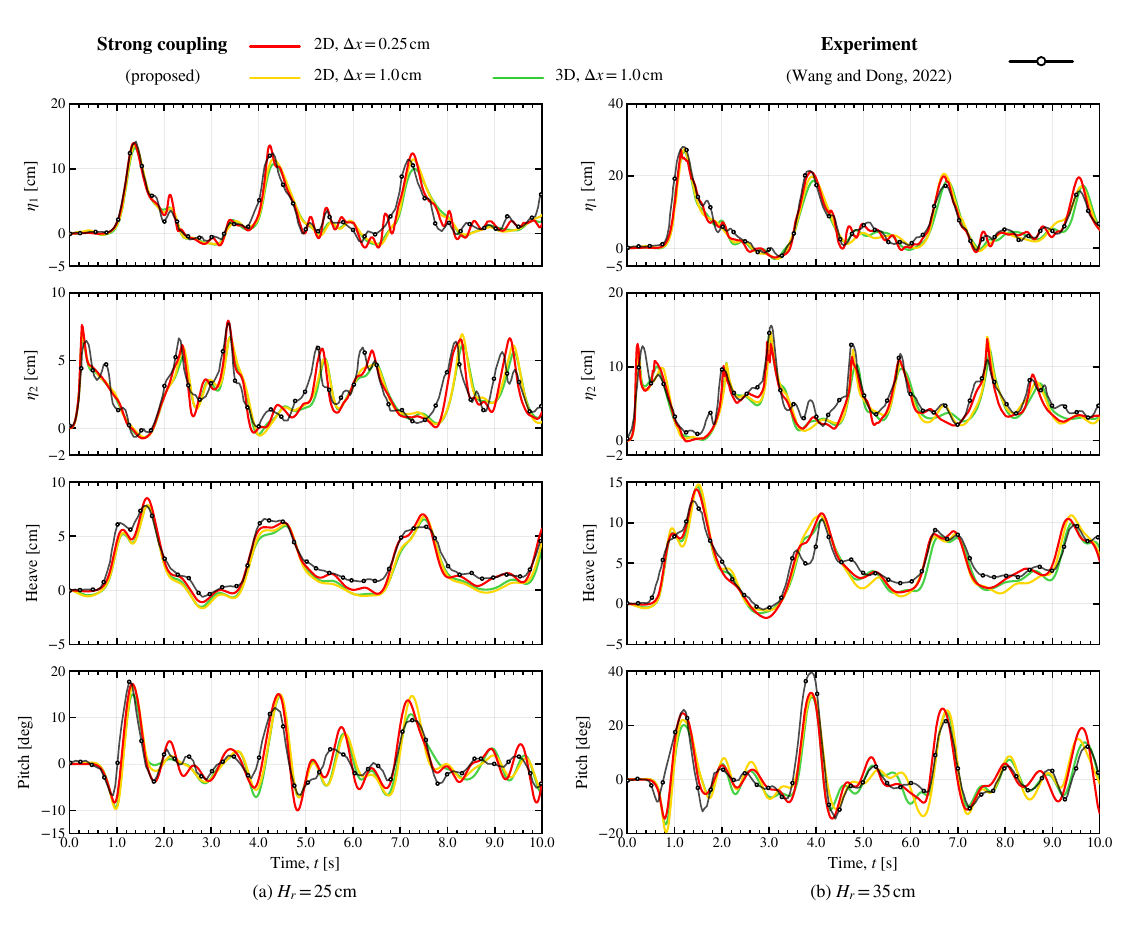}
    \caption{Heave-and-pitch responses compared with digitized experimental data from Wang and Dong [35]. (a) $H_r=25$ cm and (b) $H_r=35$ cm. Rows show wave elevations at WG1 and WG2, body heave, and body pitch. Here, $\eta_1$ and $\eta_2$ denote the wave elevations at WG1 and WG2, respectively.}
    \label{fig:dam_pitch_history}
\end{figure}

The coupling stability follows the same pattern as in the heave-only cases: the weak-coupling calculations become unstable before bore arrival, whereas the strong-coupling calculations remain stable when pitch is permitted. Figure~\ref{fig:dam_pitch_history} shows that the strong-coupling calculations reproduce the principal experimental trends in wave elevation and heave, together with the dominant oscillatory features in pitch.

\begin{figure}[tbp]
    \centering
    \includegraphics[width=0.9\linewidth]{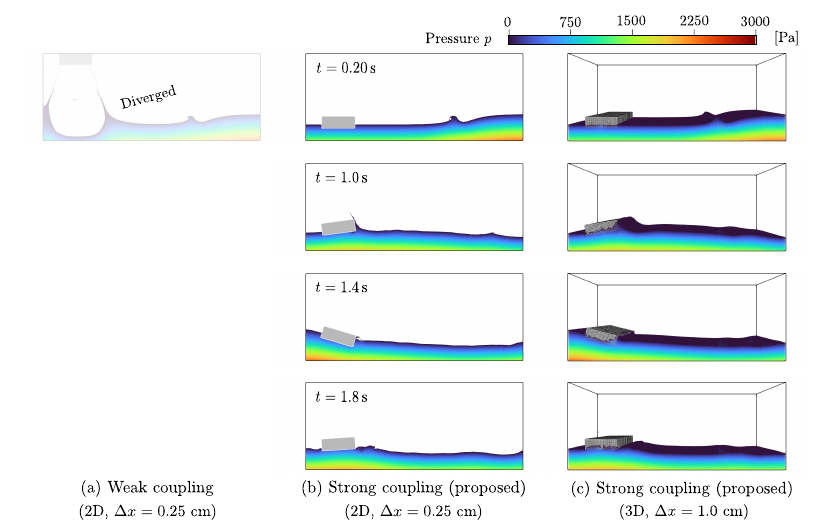}
    \caption{Heave-and-pitch snapshots for $H_r=25~\mathrm{cm}$: (a) weak coupling in 2D, (b) strong coupling in 2D, and (c) strong coupling in 3D. Rows show the indicated times; colors denote pressure in Pa. Weak-coupling results are shown only at the first snapshot.}
    \label{fig:dam_pitch_Hr25}
\end{figure}

\begin{figure}[tbp]
    \centering
    \includegraphics[width=0.9\linewidth]{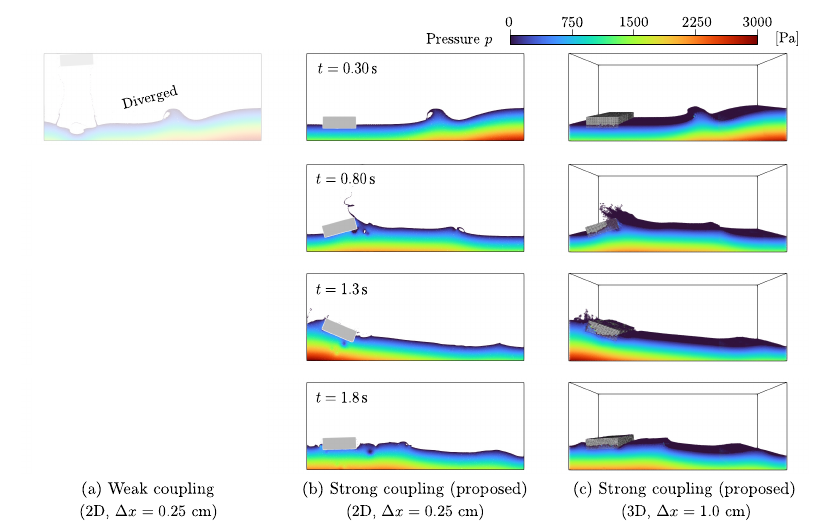}
    \caption{Heave-and-pitch snapshots for $H_r=35~\mathrm{cm}$: (a) weak coupling in 2D, (b) strong coupling in 2D, and (c) strong coupling in 3D. Rows show the indicated times; colors denote pressure in Pa. Weak-coupling results are shown only at the first snapshot.}
    \label{fig:dam_pitch_Hr35}
\end{figure}

For $H_r=25~\mathrm{cm}$, the predicted pitch follows the principal experimental trend. For $H_r=35~\mathrm{cm}$, the overall timing is reproduced, although the large positive peak near $t=4~\mathrm{s}$ is underestimated. Figures~\ref{fig:dam_pitch_Hr25} and \ref{fig:dam_pitch_Hr35} illustrate the corresponding free-surface evolution and body orientation.

Strong coupling thus remains stable under both prescribed motion constraints while capturing the principal translational and rotational responses of the floating body.

\subsection{Numerical demonstration involving multiple floating rigid bodies with complex geometries}
\label{subsec:star_float}

The preceding experimental comparisons assessed the principal wave and rigid-body responses in two and three dimensions. A two-dimensional multi-body problem is next considered to examine a different capability of the FD-SPH formulation: its applicability to complex rigid-body geometries with an evolving narrow inter-body flow.

\begin{figure}[tbp]
    \centering
    \includegraphics[width=0.95\linewidth]{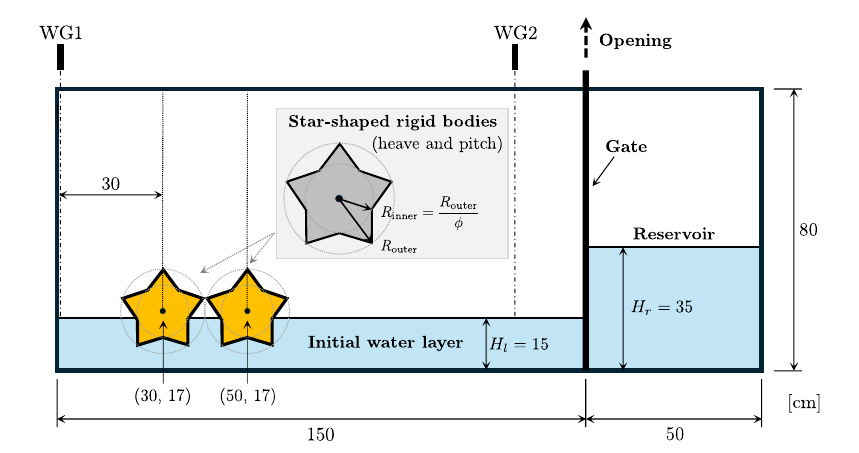}
    \caption{Wet-bed dam-break configuration with two identical star-shaped floating rigid bodies.}
    \label{fig:star_float_setup}
\end{figure}

As shown in Fig.~\ref{fig:star_float_setup}, the rectangular body used in the preceding wet-bed dam-break validation is replaced by two identical lightweight five-point star-shaped rigid bodies. The vertices of each body alternate between the outer and inner radii,
\begin{equation}
    R_{\rm inner}
    =
    \frac{R_{\rm outer}}{\varphi_g},
    \qquad
    \varphi_g=\frac{1+\sqrt{5}}{2},
\end{equation}
with $R_{\rm outer}=10~\mathrm{cm}$ and $R_{\rm inner}=6.18~\mathrm{cm}$. The rigid-body density is $\rho_s=0.20~\mathrm{g/cm^3}$, corresponding to $G_s=0.20$, and the initial body centers are located at $(30,17)~\mathrm{cm}$ and $(50,17)~\mathrm{cm}$.

The center-to-center distance is $20~\mathrm{cm}$, equal to twice the outer radius. Thus, the circumscribed circles of the two star-shaped bodies are tangent, while the star geometry leaves a narrow fluid passage between them. Both rigid bodies are allowed to heave and rotate about the out-of-plane axis, while horizontal translation is constrained. As the bodies heave and rotate, the shape and width of this passage vary continuously.

The wet-bed condition is the same as that used in the preceding validation case, with $H_r=35~\mathrm{cm}$ and $H_l=15~\mathrm{cm}$. The fluid properties and the remaining physical and numerical treatments are unchanged from the preceding two-dimensional calculation. The particle spacing and time increment are $\Delta x=0.25~\mathrm{cm}$ and $\Delta t=1.0\times10^{-4}~\mathrm{s}$, respectively, and the total simulated duration is $10~\mathrm{s}$. Anderson-accelerated strong coupling is used throughout the calculation.

Because the rigid bodies are represented within the fictitious domain, the same FD-SPH formulation is applied directly to the star-shaped bodies. Their sharp convex and re-entrant corners require neither explicit boundary reconstruction nor evaluation of surface-normal vectors, and no geometry-specific modification of the coupling or hydrodynamic-load evaluation is introduced.

\begin{figure*}[tbp]
    \centering
    \includegraphics[width=0.95\linewidth]{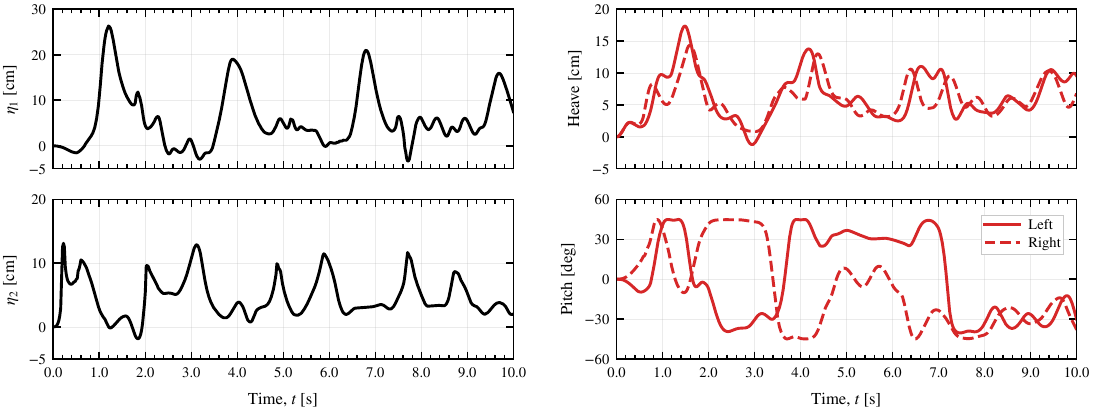}
    \caption{Time histories for the two-star wet-bed dam-break simulation: wave elevations $\eta_1$ at WG1 (upper left) and $\eta_2$ at WG2 (lower left), and heave (upper right) and pitch (lower right) of the star-shaped rigid bodies. Solid and dashed red lines indicate the left and right rigid bodies, respectively.}
    \label{fig:star_float_history}
\end{figure*}

Figure~\ref{fig:star_float_history} presents the time histories of the wave elevations at WG1 and WG2, together with the heave and pitch of the two rigid bodies. The incoming bore induces substantial heave and rotational motions. Although the bodies have identical geometry and density, their responses differ, particularly in pitch, as the local flow evolves asymmetrically.

\begin{figure*}[tbp]
    \centering
    \includegraphics[width=0.9\linewidth]{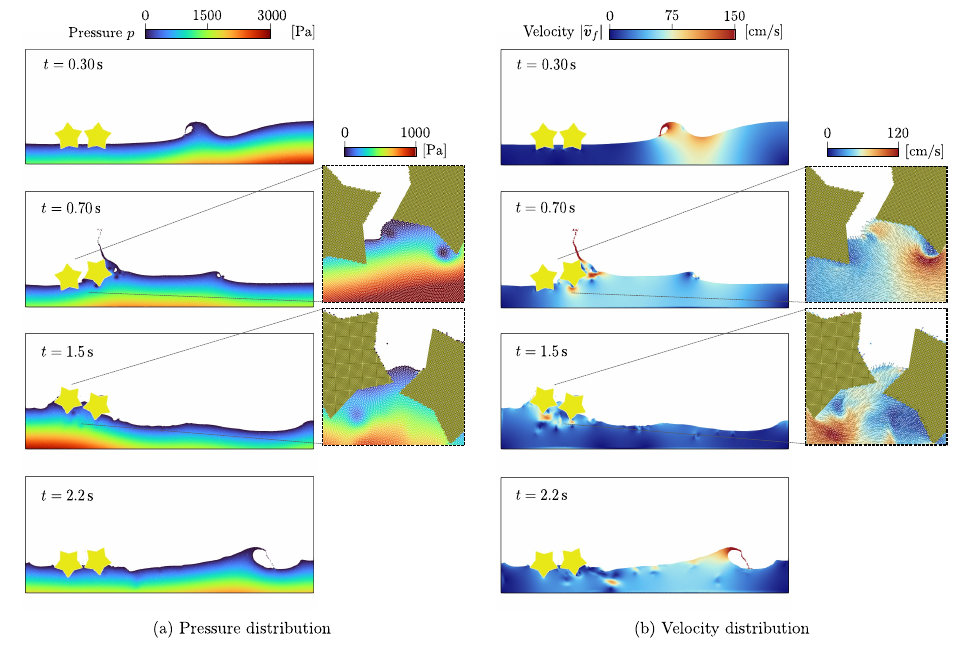}
    \caption{Snapshots of (a) pressure and (b) velocity-magnitude distributions for the two-star wet-bed dam-break simulation at $t=0.30$, $0.70$, $1.5$, and $2.2~\mathrm{s}$. Enlarged views of the inter-body region are shown at $t=0.70$ and $1.5~\mathrm{s}$. Color-bar ranges are indicated separately for the overall and enlarged views; pressure is in Pa and velocity magnitude is in cm/s.}
    \label{fig:star_float_response}
\end{figure*}

Figure~\ref{fig:star_float_response} shows the corresponding pressure and velocity-magnitude distributions at representative times. The enlarged views at $t=0.70$ and $1.5~\mathrm{s}$ illustrate the evolving narrow passage between the bodies, where locally elevated flow speeds develop as the bodies move and rotate. These snapshots also demonstrate that the pressure and velocity fields can be obtained for the complex moving geometries without explicit surface reconstruction or geometry-specific boundary treatment.

The coupling iteration converged at all physical time steps before reaching $K_{\max}=2000$, requiring 23.9 iterations on average, with a 95th-percentile iteration count of 34.

These results demonstrate stable partitioned coupling for two geometrically complex rigid bodies under transient free-surface flow, including the evolving narrow passage between them.

\section{Conclusions}
\label{sec:conclusions}

This study developed and assessed a rigid-body-specific implementation of partitioned iterative fictitious-domain smoothed particle hydrodynamics (FD-SPH) for free-surface fluid--rigid-body interaction, building on the strong-coupling framework previously established for deformable structures. Hydrodynamic force and torque are approximated by volume integration of the extended-fluid stress divergence and used to update rigid-body translation and rotation through the Newton--Euler equations. Within each physical time step, the kinematic mismatch between the pressure-projected fictitious-fluid and rigid-body velocities is returned to the fluid solver, allowing the pressure field and rigid-body response to be recomputed through partitioned iterations while retaining separate solvers.

The two-dimensional settling-disk benchmark showed agreement with published numerical reference results, with nearly identical responses from single-pass and iterative coupling. For buoyancy-driven cubes, single-pass coupling became unstable at $G_s=0.1$ and $0.2$, whereas iterative strong coupling remained stable over the tested range $G_s=0.1$--$1.0$. At $G_s=0.1$, reducing the time increment by a factor of up to 500 did not prevent the single-pass instability. The 45~s heave-only simulations also provided detailed full-run coupling-iteration statistics. The predicted equilibrium heights and observed three-dimensional orientations were consistent with hydrostatic expectations and previously reported stable configurations, respectively. These results demonstrate the benefit of within-step residual feedback for the examined lightweight-body conditions.

Comparisons with wet-bed dam-break experiments showed that the strongly coupled formulation captures the principal features of the free-surface elevation and floating-body heave and pitch responses in two and three dimensions. A simulation with two star-shaped rigid bodies further demonstrated geometric applicability to multiple moving bodies with sharp convex and re-entrant corners and an evolving narrow inter-body flow, without explicit interface reconstruction or surface-normal evaluation. This multi-body example is a numerical demonstration rather than an independent experimental validation, and quantitative errors in under-resolved narrow gaps have not been established by a spatial-refinement study.

From a fractional-step perspective, PMS and single-pass FD-SPH share a momentum-based treatment of rigid-body motion. The explicitly stated linear and angular momentum conditions show how the PMS momentum-preserving projection reproduces the Newton--Euler translational and rotational updates, including the gyroscopic contribution, under consistent discrete momentum transfer. The present iterative FD-SPH formulation further incorporates within-step residual feedback to reconcile the fluid and rigid-body solutions. Across all reported production strong-coupling simulations, the partitioned iterations satisfied the prescribed relative-residual criterion at every physical time step before reaching $K_{\max}$. Together, the results support the use of the proposed formulation for the tested lightweight floating-body problems, reproducing the principal fluid and rigid-body responses and accommodating complex moving geometries in free-surface flows.

\appendix


\section{SPH spatial discretization and particle shifting}
\label{subsec:SPH_discretization}

The governing equations presented in Section~\ref{sec:gov_eqs} are spatially discretized using the smoothed particle hydrodynamics (SPH) method~\cite{Lucy_SPH_1977,Gingold-Monaghan_SPH_1977}. Within the fictitious-domain formulation, both the physical-fluid domain $\Omega_f$ and the rigid-body domain $\Omega_s$ are represented by SPH particles, and the same differential operators are applied over the extended domain $\widetilde{\Omega}_f=\Omega_f\cup\Omega_s$. To retain spatial accuracy under irregular particle distributions, the second-order consistent SPH formulation, referred to as SPH(2), is employed~\cite{Asai_SPH2_2023,Tsuji_SPH2_DPS_seepage_2025}. The particle-shifting procedure used in the ALE transport velocity is also summarized in this appendix.


\subsection{Kernel approximation and corrected kernel gradient}
\label{subsec:sph_kernel}

Let $\phi(\bm x,t)$ denote a scalar field defined over the extended computational domain. Its SPH kernel approximation at particle $i$ is expressed as
\begin{equation}
\label{eq:sph_integral_kernel}
\phi(\bm x_i,t) \approx \int_{\widetilde{\Omega}_f} \phi(\bm x',t) w(|\bm x'-\bm x_i|,h)\,\mathrm{d}V(\bm x'),
\end{equation}
where $\bm x'$ denotes the integration position. For the subsequent particle discretization, $\bm r_{ij}=\bm x_j-\bm x_i$ and $r_{ij}=|\bm r_{ij}|$; $h$ is the smoothing length, and $w$ is the smoothing kernel.

The cubic spline kernel is employed in the present study,
\begin{equation}
\label{eq:cubic_spline_kernel}
w(r_{ij},h) = w_{ij} = \alpha_d^c
\begin{cases}
\displaystyle 1-\frac{3}{2}\left(\frac{r_{ij}}{h}\right)^2+\frac{3}{4}\left(\frac{r_{ij}}{h}\right)^3, & 0\le\displaystyle\frac{r_{ij}}{h}\le1, \\[3mm]
\displaystyle \frac{1}{4}\left(2-\frac{r_{ij}}{h}\right)^3, & 1<\displaystyle\frac{r_{ij}}{h}\le2, \\[3mm]
0, & \text{otherwise},
\end{cases}
\end{equation}
where
\begin{equation}
\alpha_d^c =
\begin{cases}
\displaystyle \frac{10}{7\pi h^2}, & d=2, \\[3mm]
\displaystyle \frac{1}{\pi h^3}, & d=3.
\end{cases}
\end{equation}
Here, $\alpha_d^c$ is the cubic-spline normalization coefficient, and $d\in\{2,3\}$ denotes the spatial dimension. The influence radius is set to $r_e=2h=2.4\Delta x$, where $\Delta x$ denotes the initial particle spacing.

For a target particle $i$, the neighboring-particle set is defined as
\begin{equation}
\label{eq:SPH_neighbor_set}
\mathbb S_i = \left\{j\in\mathcal P \;\middle|\; j\neq i,\; r_{ij}<r_e\right\},
\end{equation}
where $\mathcal P=\mathcal P_f\cup\mathcal P_s$ is the unified particle set introduced in Section~\ref{subsec:ale_governing}. Thus, no distinction is made between physical-fluid and fictitious-fluid particles when evaluating the SPH differential operators.

Using the representative particle volume $V_j$, the SPH interpolation of a scalar field is written as
\begin{equation}
\label{eq:SPH_phi}
\langle\phi\rangle_i = \sum_{j\in\mathbb S_i\cup\{i\}} V_j\phi_jw_{ij}.
\end{equation}

Throughout this appendix, the kernel gradient is taken with respect to the target-particle position $\bm x_i$:
\begin{equation}
\nabla w_{ij} \equiv \nabla_{\bm x_i}w(r_{ij},h) = -\frac{\partial w(r_{ij},h)}{\partial r_{ij}}\frac{\bm r_{ij}}{r_{ij}}, \qquad j\ne i.
\end{equation}
This convention is consistent with $\bm r_{ij}=\bm x_j-\bm x_i$ and fixes the sign of the corrected gradient used below.

Gradient correction restores the exact reproduction of linear-field gradients under irregular particle distributions~\cite{Bonet_GC_SPH_1999}, provided that the local moment matrix is nonsingular. Following the notation used in the SPH(2) formulation~\cite{Asai_SPH2_2023,Tsuji_SPH2_DPS_seepage_2025}, the corrected kernel gradient is defined as
\begin{equation}
\label{eq:corrected_kernel_gradient}
\widetilde{\nabla}w_{ij}=\mathbf L_i\nabla w_{ij},
\qquad \mathbf L_i=\left[\sum_{j\in\mathbb S_i}V_j(\nabla w_{ij}\otimes\bm r_{ij})\right]^{-1}.
\end{equation}
The corresponding first-order corrected gradient is
\begin{equation}
\label{eq:SPH_grad_1st}
\left\langle\nabla\phi\right\rangle_i^{(1)} = \sum_{j\in\mathbb S_i} V_j\phi_{ij}\widetilde{\nabla}w_{ij},
\end{equation}
where $\phi_{ij}=\phi_j-\phi_i$.

\subsection{Second-order SPH formulation}
\label{subsec:sph2_formulation}

The SPH(2) formulation evaluates first- and second-order spatial derivatives in a decoupled manner based on the second-order Taylor expansion~\cite{Asai_SPH2_2023}. The formulation is presented below in three-dimensional form; the corresponding two-dimensional form is obtained by removing the components associated with the $z$ direction.

For a scalar field $\phi$, define the vector of second derivatives as
\begin{equation}
\label{eq:SPH2_second_derivative_vector}
\bm d_i^{(2)} =
\begin{bmatrix}
\left\langle \dfrac{\partial^2\phi}{\partial x^2} \right\rangle_i^{(2)} &
\left\langle \dfrac{\partial^2\phi}{\partial y^2} \right\rangle_i^{(2)} &
\left\langle \dfrac{\partial^2\phi}{\partial z^2} \right\rangle_i^{(2)} &
2\left\langle \dfrac{\partial^2\phi}{\partial x\partial y} \right\rangle_i^{(2)} &
2\left\langle \dfrac{\partial^2\phi}{\partial y\partial z} \right\rangle_i^{(2)} &
2\left\langle \dfrac{\partial^2\phi}{\partial z\partial x} \right\rangle_i^{(2)}
\end{bmatrix}^{\top}.
\end{equation}

The Full-Inverse SPH(2) formulation employs the coefficient
\begin{equation}
\label{eq:Fcoef_3D}
F_{ij}
=
\frac{\bm r_{ij}\cdot\widetilde{\nabla}w_{ij}}
{|\bm r_{ij}|^4}.
\end{equation}
The correction vectors are defined as
\begin{equation}
\label{eq:C_alpha_beta_def}
\bm B_{\alpha\beta,i}
=
\sum_{k\in\mathbb S_i}
V_k r_{ik}^{\alpha}r_{ik}^{\beta}
\widetilde{\nabla}w_{ik},
\qquad \alpha,\beta\in\{x,y,z\},
\end{equation}
where $r_{ik}^{\alpha}$ denotes the $\alpha$-component
of the relative position vector $\bm r_{ik}$.

The vectors $\bm q_{ij}$ and $\bm p_{ij}$ are constructed
from the quadratic monomials of
$\bm r_{ij}=(x_{ij},y_{ij},z_{ij})^\top$ as
\begin{equation}
\label{eq:pq_vec_3D}
\bm q_{ij}
=
\begin{bmatrix}
x_{ij}^2\\
y_{ij}^2\\
z_{ij}^2\\
x_{ij}y_{ij}\\
y_{ij}z_{ij}\\
z_{ij}x_{ij}
\end{bmatrix},
\qquad
\bm p_{ij}
=
\begin{bmatrix}
x_{ij}^2-\bm r_{ij}\cdot\bm B_{xx,i}\\
y_{ij}^2-\bm r_{ij}\cdot\bm B_{yy,i}\\
z_{ij}^2-\bm r_{ij}\cdot\bm B_{zz,i}\\
x_{ij}y_{ij}-\bm r_{ij}\cdot\bm B_{xy,i}\\
y_{ij}z_{ij}-\bm r_{ij}\cdot\bm B_{yz,i}\\
z_{ij}x_{ij}-\bm r_{ij}\cdot\bm B_{zx,i}
\end{bmatrix}.
\end{equation}

Using these quantities, the second-order moment matrix is
defined as
\begin{equation}
\label{eq:moment_matrix_3D}
\mathbf M_i
=
\sum_{j\in\mathbb S_i}
V_j F_{ij}\bm q_{ij}\bm p_{ij}^{\top}.
\end{equation}

To mitigate ill-conditioning caused by incomplete particle
supports near the free surface, a localized diagonal
regularization is introduced following Matsunaga and
Koshizuka~\cite{Matsunaga_StabLSMPS_2022}.
The regularized moment matrix is defined as
\begin{equation}
\label{eq:moment_matrix_regularization}
\mathbf M_i^{\mathrm{reg}}
=
\begin{cases}
\mathbf M_i+\alpha\bm 1,
& i\in\mathcal P_{\mathrm{FS}},\\
\mathbf M_i,
& i\notin\mathcal P_{\mathrm{FS}}.
\end{cases}
\end{equation}
Here, $\mathcal P_{\mathrm{FS}}\subset\mathcal P$
denotes the free-surface particle set identified over
the extended fluid domain, including atmosphere-exposed
fictitious-fluid particles belonging to the rigid bodies.
The multistage detection procedure is based on
Marrone et al.~\cite{Marrone_SPH-FSD_2010}
and Matsunaga and
Koshizuka~\cite{Matsunaga_StabLSMPS_2022}.
The regularization parameter is empirically set to
$\alpha=0.2$, and $\bm 1$ denotes the identity matrix.
The same free-surface particle set is used to impose
the zero-gauge-pressure Dirichlet condition in
Eq.~\eqref{eq:free_surface_pressure_bc}.

Using the regularized moment matrix, the Full-Inverse
SPH(2) approximation of the second derivatives is
finally expressed as
\begin{equation}
\label{eq:SPH2_2nd_drv_3D}
\bm d_i^{(2)}
=
2(\mathbf M_i^{\mathrm{reg}})^{-1}
\sum_{j\in\mathbb S_i}
V_jF_{ij}\bm q_{ij}
\left(
\phi_{ij}
-
\bm r_{ij}\cdot
\left\langle\nabla\phi\right\rangle_i^{(1)}
\right).
\end{equation}

Once the second derivatives have been evaluated from Eq.~\eqref{eq:SPH2_2nd_drv_3D}, the second-order gradient is obtained as
\begin{equation}
\label{eq:SPH_grad_2nd_3D}
\left\langle \nabla\phi \right\rangle_i^{(2)} = \sum_{j\in\mathbb S_i} V_j \left[ \phi_{ij} - \frac{1}{2} \sum_{I=1}^{d} \sum_{J=1}^{d} r_{ij}^{I} r_{ij}^{J} \left\langle \frac{ \partial^2\phi }{ \partial x^{I}\partial x^{J} } \right\rangle_i^{(2)} \right] \widetilde{\nabla}w_{ij}.
\end{equation}

Here, $x^I$ denotes the $I$-th Cartesian coordinate, whereas $r_{ij}^I=x_j^I-x_i^I$ is the corresponding component of the relative position vector $\bm r_{ij}$. For a vector field $\bm\phi$, the divergence is evaluated from the corresponding component-wise second-order gradients as
\begin{equation}
\label{eq:SPH_div_2nd_3D}
\left\langle \nabla\cdot\bm\phi \right\rangle_i^{(2)} = \sum_{I=1}^{d} \left\langle \frac{ \partial\phi_I }{ \partial x^I } \right\rangle_i^{(2)}.
\end{equation}
The scalar Laplacian is obtained as the trace of the second-derivative tensor,
\begin{equation}
\label{eq:SPH_lap_2nd_3D}
\left\langle \nabla^2\phi \right\rangle_i^{(2)} = \sum_{I=1}^{d} \left\langle \frac{ \partial^2\phi }{ (\partial x^I)^2 } \right\rangle_i^{(2)}.
\end{equation}
For a vector field, the Laplacian is evaluated component-wise.

The notation $\langle\cdot\rangle^{(2)}$ denotes the SPH(2) approximation constructed from the second-order Taylor expansion. In the present formulation, these operators are used for the spatial discretization of the pressure gradient, velocity divergence, advection, and Laplacian terms appearing in the ALE-form extended governing equations and the pressure Poisson equation.


\subsection{Particle shifting in the ALE description}
\label{subsec:particle_shifting}

Particle shifting is introduced for the physical-fluid particles to maintain a regular particle distribution during the ALE calculation. The procedure follows Tsuji et al.~\cite{Tsuji_SPH2_DPS_seepage_2025}, in which the total shifting displacement is composed of a particle-homogenization component based on the optimized particle shifting (OPS) of Khayyer et al.~\cite{Khayyer_OPS_2017} and a density-based component for suppressing numerical density fluctuations,
\begin{equation}
\label{eq:particle_shift_total}
\delta\bm r_i
=
\delta\bm r_i^{\rm homo}
+
\delta\bm r_i^{\rm vol},
\qquad i\in\mathcal P_f .
\end{equation}
The particle-homogenization and density-based shifting formulations, including their free-surface treatment and displacement limits, are detailed in Tsuji et al.~\cite{Tsuji_SPH2_DPS_seepage_2025}. For the present pure-fluid calculation, their density-based formulation is specialized by setting the porosity parameter to $\epsilon_i=1$, so that the reference numerical density reduces to the constant fluid density $\rho_f$. The numerical-density evaluation and numerical-pressure function used for this shifting component are described in the cited formulation.

The resulting shifting displacement is incorporated into the ALE transport velocity of the physical-fluid particles as
\begin{equation}
\label{eq:shift_velocity}
\bm w_i
=
\widetilde{\bm v}_{f,i}
+
\frac{\delta\bm r_i}{\Delta t},
\qquad i\in\mathcal P_f ,
\end{equation}
consistent with the ALE formulation introduced in Section~\ref{subsec:ale_governing}. No independent particle shifting is applied inside the rigid-body domain. Instead, the particles in $\mathcal P_s$ are transported according to the rigid-body motion, $\bm w_i=\bm v_{s,i}$ for $i\in\mathcal P_s$.


\section*{CRediT authorship contribution statement}

\textbf{Kumpei Tsuji:} Conceptualization, Methodology, Software, Validation, Formal analysis, Investigation, Data curation, Visualization, Funding acquisition, Writing -- original draft, Writing -- review \& editing, Project administration.
\textbf{Ciara Galvin:} Software, Investigation, Validation, Visualization, Writing -- review \& editing.
\textbf{Hiroyuki Omura:} Conceptualization, Methodology, Writing -- review \& editing.
\textbf{Naoto Mitsume:} Conceptualization, Methodology, Supervision, Writing -- review \& editing.
\textbf{Mitsuteru Asai:} Conceptualization, Methodology, Supervision, Funding acquisition, Writing -- review \& editing.

\section*{Declaration of competing interest}

The authors declare that they have no known competing financial interests or personal relationships that could have appeared to influence the work reported in this paper.

\section*{Acknowledgements}

The computations in this study were carried out using the computer resources offered under the category of General Projects by the Research Institute for Information Technology, Kyushu University.

\section*{Funding}

This work was supported by JSPS KAKENHI Grant Numbers JP24K17481, JP23K26356, and JP23KK0182, and by the River Fund of The River Foundation, Japan. The funding bodies had no role in the study design, data collection, analysis, interpretation, manuscript preparation, or decision to submit the article for publication.

\section*{Data availability}

The data supporting the findings of this study are available from the corresponding author upon reasonable request.

\section*{Declaration of generative AI and AI-assisted technologies in the manuscript preparation process}

During the preparation of this work, the authors used OpenAI ChatGPT to assist with language editing, manuscript clarity, organization, and pre-submission consistency checks. The authors critically evaluated and independently verified the suggestions, revised the manuscript as appropriate, and take full responsibility for the content of the published article.

\bibliographystyle{elsarticle-num}
\bibliography{references}

\end{document}